\documentclass[11pt]{article}

\usepackage{amsthm,amsmath,amssymb,bbm}
\usepackage{natbib}
\usepackage{multirow}
\usepackage[pdftex]{graphicx}
\usepackage{subfigure}

\usepackage{setspace}
\usepackage{makecell}
\usepackage{booktabs}
\usepackage{array}
\usepackage{tabularx}
\usepackage{tabulary}
\usepackage{caption}
\usepackage{booktabs}
\usepackage{pgfplots}
\usepackage{url}
\usepackage{algorithm}
\usepackage{algorithmic}
\usepackage{bm}
\usepackage{wrapfig}
\usepackage{lipsum}
\usepackage{mathrsfs}
\usepackage{dsfont}
\usepackage{titling}
\usepackage{txfonts}

\usepackage{xr}

\usepackage[colorlinks,
linkcolor=red,
anchorcolor=blue,
citecolor=blue
]{hyperref}

\usepackage{smile}

\def\##1\#{\begin{align}#1\end{align}}
\def\$#1\${\begin{align*}#1\end{align*}}

\usepackage{relsize}

\newcommand{\bfsym}[1]{\ensuremath{\boldsymbol{#1}}}
       \def \bbeta    {\bfsym{\beta}}

\newcommand{\Rom}[1]{\text{\uppercase\expandafter{\romannumeral #1\relax}}}

\usepackage{geometry}
\usepackage{enumitem}

\numberwithin{equation}{section}

\title{A Unified Algorithm for Penalized Convolution Smoothed Quantile Regression}

\author{Rebeka Man\thanks{Department of Statistics, University of Michigan, Ann Arbor, Michigan 48109, USA. E-mail:\href{mailto:mrebeka@umich.edu}{\textsf{mrebeka@umich.edu}}.},~~Xiaoou Pan\thanks{Department of Mathematics, University of California, San Diego, La Jolla, CA 92093, USA. E-mail:\href{mailto:xip024@ucsd.edu}{\textsf{xip024@ucsd.edu}}.},~~Kean Ming Tan\thanks{Department of Statistics, University of Michigan, Ann Arbor, Michigan 48109, USA. E-mail:\href{mailto:keanming@umich.edu}{\textsf{keanming@umich.edu}}.},~~and~Wen-Xin Zhou\thanks{Department of Mathematics, University of California, San Diego, La Jolla, CA 92093, USA. E-mail:\href{mailto:wez243@ucsd.edu}{\textsf{wez243@ucsd.edu}}.} }

\date{}

\begin{document}
\maketitle

\vspace{-.8in}
\begin{abstract}
Penalized quantile regression (QR) is widely used for studying the relationship between a response variable and a set of predictors under data heterogeneity in high-dimensional settings. Compared to penalized least squares,  scalable algorithms for fitting penalized QR  are lacking  due to the non-differentiable piecewise linear loss function. To overcome the lack of smoothness,  a recently proposed convolution-type smoothed method brings an interesting tradeoff between statistical accuracy and computational efficiency for both standard and penalized quantile regressions. In this paper, we propose a unified algorithm for fitting penalized convolution smoothed quantile regression with various commonly used convex penalties, accompanied by an \texttt{R}-language package \texttt{conquer} available from the Comprehensive R Archive Network. We perform extensive numerical studies to demonstrate the superior performance of the proposed algorithm over existing methods in both statistical and computational aspects.   
We  further exemplify the proposed algorithm by fitting a fused lasso additive QR model on the world happiness data.   
\end{abstract}

\section{Introduction}
\label{sec:1}
Let $y \in \mathbb{R}$ be a scalar response variable of interest,  and $\bx \in \mathbb{R}^p$ be a $p$-dimensional vector of covariates. Since the seminal work of \citet{KB1978},  quantile regression (QR) has become an indispensable tool for understanding pathways of dependence between $y$ and $\bx$, which is irretrievable through conditional mean regression analysis via the least squares method.  Motivated by a wide range of applications,  various models have been proposed and studied for QR,  from parametric to nonparametric and from low- to high-dimensional
covariates.
We refer the reader to \citet{K2005} and \citet{KCHP2017} for a comprehensive exposition of quantile regression.

Consider a  high-dimensional linear QR model in which the number of covariates, $p$, is larger than the number of observations, $n$.   In this setting,  different low-dimensional structures have been imposed on the regression coefficients, thus motivating the use of various penalty functions.
One of the most widely used assumption is the sparsity, which assumes that only a small number of predictors are associated with the response.  Quantile regression models are capable of capturing heterogeneity in the set of important predictors at different quantile levels of the response distribution caused by,  for instance,  heteroscedastic variance.
For fitting sparse models in general,  various sparsity-inducing penalties have been introduced,  such as the lasso ($\ell_1$-penalty)  \citep{Tibs1996},  elastic net (hybrid  of $\ell_1$/$\ell_2$-penalty) \citep{ZH2005}, smoothly clipped absolute deviation (SCAD) penalty  \citep{FL2001} and minimax concave (MC) penalty \citep{ZMCP2010}.
Other commonly used regularizers that induce different types of structures include the group lasso \citep{YL2006}, sparse group lasso \citep{SFHT2013} and fused lasso \citep{TSRZK2005}, among others. We refer to the monographs \citet{BVG11}, \citet{HTW2015} and \citet{W2019} for systematic introductions of high-dimensional statistical learning.

The above penalties/regularizers have been  extensively studied when applied with the least squares method,  accompanied with the user-friendly and efficient software \texttt{glmnet} \citep{glmnet}.
Quantile regression, on the other hand, involves minimizing a non-differentiable piecewise linear loss,  known as the check function. Although a handful of algorithms have been developed based on either linear programming or  the alternating direction method of multipliers (ADMM) \citep{KN2005, LZ2008, PW2015, Yu2017,Gu2018},   there is not much software that is nearly as efficient as \texttt{glmnet} available for penalized quantile regressions.
As the most recent progress,  \citet{YH2017} proposed a semismooth Newton coordinate descent (SNCD) algorithm for solving a Huberized  QR  with the elastic net penalty; \citet{Yu2017} and  \cite{Gu2018} respectively  proposed  ADMM-based algorithms for solving folded-concave penalized QR. See \cite{Gu2018}  for a detailed computational comparison between the SNCD and ADMM-based algorithms, which favors the latter.
The primary computation effort of each ADMM update  is to evaluate the
inverse of a $p\times p$ or $n \times  n$ matrix. This can be computationally expensive  when both $n$ and $p$ are large. Compared to sparsity-inducing regularization,  the computational development of penalized QR with other important penalties, such as the group lasso,  is still scarce.  For instance,  quantile regression with the group lasso penalty can be formulated as a second-order cone programming (SOCP) problem \citep{KK2011}, solvable by general-purpose optimization toolboxes. These toolboxes are only adapted to small-scale problems and usually lead to solution with very high precision (low duality gap).  For large-scale datasets, they tend to be too slow, or often run out of memory.

To resolve the non-differentiability issue of the check loss, \cite{H1998} proposed to smooth the check loss directly using a kernel function. This approach, however,  leads to a non-convex loss function which brings further computational issues especially in high dimensions.  Recently,  \cite{FGH2019} employed a convolution-type smoothing technique to introduce the smooth quantile regression (SQR) without sacrificing convexity.   Convolution smoothing turns the non-differentiable check function
into a twice-differentiable,  convex and locally strongly convex surrogate,  which admits fast and scalable gradient-based algorithms to perform optimization \citep{HPTZ2020}.  Theoretically, the SQR estimator is asymptotically first-order equivalent to the standard QR estimator, and enjoys desirable statistical properties  \citep{FGH2019, HPTZ2020}.   For high-dimensional sparse models,  \citet{TWZ2021} proposed an iteratively reweighted $\ell_1$-penalized SQR estimator that achieves oracle rate of convergence when the signals are sufficiently strong. They also proposed  coordinate descent and ADMM-based algorithms for (weighted) $\ell_1$-penalized SQR with the uniform and Gaussian kernels. These algorithms, however,  do not adapt to more general kernel functions as well as penalties.

In this paper,  we introduce a major variant of the local adaptive majorize-minimization (LAMM) algorithm \citep{FLSF2018} for fitting convolution smoothed quantile regression that applies to any kernel function and a wide range of convex penalties. The main idea  is to construct an isotropic quadratic objective function  that locally majorizes the smoothed quantile loss such that closed-form updates are available at each iteration.  The quadratic coefficient is adaptively chosen in order to guarantee the decrease of the objective function. In a sense,  the LAMM algorithm can be viewed as a generalization of the iterative shrinkage-thresholding algorithm (ISTA) \citep{BT2009}.
Compared to the interior point methods (for solving linear programming and SOCP problems) as well as ADMM,  LAMM is a simpler gradient-based algorithm that is particularly suited  for large-scale problems,  where the dominant computational effort is a relatively cheap matrix-vector multiplication at each step.   The (local) strong convexity of the convolution smoothed quantile loss facilitates the convergence of such a first order method.  A key advantage of the proposed algorithm over those in \citet{TWZ2021} is that  it can be applied to a broad class of convex penalties, typified by the lasso, elastic net, group lasso and sparse group lasso,  and to any continuous kernel function. The proposed algorithm is implemented in the \texttt{R} package \texttt{conquer} \citep{HPTZ2022} for fitting penalized (smoothed) quantile regression, including all the penalties considered in this paper.

The remainder of the paper is organized as follows.  Section~\ref{sec:bkgrd} briefly revisits quantile regression and its convolution smoothed counterpart.  In Section~\ref{sec:psalg},  we  describe a general  local adaptive majorize-minimization principal for solving penalized smoothed quantile regression with four types of convex penalties, which are the lasso ($\ell_1$-penalty),  elastic net (hybrid of $\ell_1$/$\ell_2$-penalty),  group lasso (weighted $\ell_2$-penalty) and sparse group lasso (hybrid of $\ell_1$- and weighted $\ell_2$-penalty).
The computational and statistical efficiency of the proposed algorithm is demonstrated
via extensive simulation studies in Section~\ref{sec:ns}.
In Section~\ref{sec:5}, we further exemplify the the proposed algorithm by fitting a fused lasso additive QR model on the world happiness data.


\section{Quantile Regression and Convolution Smoothing}
\label{sec:bkgrd}
Let ${\bx} \in \mathbb{R}^p$ be a $p$-dimensional covariates and let $y \in \mathbb{R}$ be a scalar response variable.
Given a quantile level $\tau \in (0,1)$ of interest,  assume that the $\tau$-th conditional quantile of $y$ given $\bx$  follows a linear model $F^{-1}_{y|\bx}(\tau)=\bx^T\bbeta^*(\tau)$, where $\bbeta^*(\tau) = \{  \beta_1^*(\tau),\ldots, \beta_p^*(\tau) \}^T \in \mathbb{R}^{p}$.
For notational convenience, we set $x_1 \equiv 1$ such that $\beta_1^*$ is the intercept.   
Moreover, we suppress the dependency of $\bbeta^*(\tau)$ on $\tau$ throughout the paper.  Let $\{y_i,\bx_i\}^n_{i=1}$ be a random sample of size $n$ from $(y,\bx)$.  The standard quantile regression estimator is defined as the solution to the optimization problem \citep{KB1978}
\begin{equation}
\label{eq:qroptim}
\displaystyle{\minimize_{\bbeta \in \mathbb{R}^{p}} ~\frac{1}{n} \sum_{i=1}^n \rho_\tau(y_i-\bx_i^T\bbeta)},
\end{equation}
where $\rho_\tau(u)=u\{\tau-\mathbbm{1}(u<0)\}$ is the quantile loss, also known as the check function.
Although the quantile loss is convex,  its non-differentiability (even only at one point)  prevents gradient-based algorithms to be efficient.  In this case,  subgradient methods typically exhibit very slow (sublinear) convergence and hence are not computationally stable.   A more widely acknowledged approach is to formulate the optimization problem \eqref{eq:qroptim} as a linear program,
solvable by the simplex algorithm or interior point methods. The latter has an average-case computational complexity that grows as a cubic function of $p$ \citep{PK1997}.

For fitting a sparse QR model in high dimensions,  a natural parallel to the lasso \citep{Tibs1996} is the $\ell_1$-penalized QR (QR-lasso) estimator \citep{BC2011},  defined as a solution to the optimization problem
\begin{equation}
\label{eq:l1penalized}
\displaystyle{\minimize_{\bbeta \in \mathbb{R}^{p}}~\frac{1}{n} \sum_{i=1}^n \rho_\tau(y_i-\bx_i^T\bbeta)}+ \lambda \|\bbeta\|_1,
\end{equation}
where $\lambda >0$ is a regularization parameter that controls (indirectly) the sparsity level of the solution, and $\|\cdot\|_1$ denotes the $\ell_1$-norm.  We refer to \cite{WWL2012} and \cite{ZPH2015} for further extensions to adaptive $\ell_1$ and folded concave penalties. Computationally,  all of these sparsity-driven penalized methods boil down to  solving a weighted $\ell_1$-penalized QR loss minimization problem that is of the form
\begin{equation} \label{eq:weightedl1} 
\displaystyle{\minimize_{\bbeta \in \mathbb{R}^{p}}~\frac{1}{n} \sum_{i=1}^n \rho_\tau(y_i-\bx_i^T\bbeta)}+ \sum_{j=1}^p \lambda_j | \beta_j | ,
\end{equation}
where $\lambda_j \geq 0$ for each $1\leq j\leq p$. Thus far,  the most notable methods for solving \eqref{eq:l1penalized} or more generally \eqref{eq:weightedl1} include linear programming \citep{K2022},  coordinate descent algorithms \citep{PW2015,YH2017} and  ADMM-based algorithms \citep{Yu2017, Gu2018}.  Among these,  ADMM-based algorithms have the best overall performance as documented in \cite{Gu2018}.  The dominant computational effort  of each ADMM update is the inversion of a $p\times p$ or an $n \times  n$ matrix.   For genomics data that typically has a small sample size, say in the order of hundreds,  ADMM works well even when the dimension $p$ is in the order of thousands or tens of thousands.
However, there is not much efficient algorithm,  that is also scalable to $n$, available for (weighted) $\ell_1$-penalized QR, let alone for more general penalties.

To address the non-differentiability of the quantile loss, \citet{FGH2019}  proposed a convolution smoothed approach to quantile regression, resulting in a twice-differentiable, convex and (locally) strongly convex loss function.  On the statistical aspect,   \citet{FGH2019} and \citet{HPTZ2020}  established the asymptotic and finite-sample properties for the smoothed QR (SQR) estimator, respectively.    \citet{TWZ2021} further considered the penalized SQR with iteratively reweighted $\ell_1$-regularization and proved oracle properties under a minimum signal strength condition. 
Let  $K(\cdot)$  be a symmetric, non-negative kernel function, that is, $K(u) = K(-u) \geq 0$ for any $u\in \RR$, and $\int_{-\infty}^\infty K(u) {\rm d} u = 1$.
Given a bandwidth parameter $h>0$,  the $\ell_1$-penalized SQR (SQR-lasso) estimator is defined as the solution to 
\begin{equation}\label{sqrloss}
\displaystyle{\minimize_{\bbeta \in \mathbb{R}^{p}}~\frac{1}{n} \sum_{i=1}^n \ell_{h,\tau}(y_i-\bx_i^T\bbeta) + \lambda\|\bbeta\|_1, ~~\text{where}~~ \ell_{h,\tau}(u) = \frac{1}{h} \int^\infty_{-\infty} \rho_\tau(v)K\left(\frac{v-u}{h}\right)\text{d}v}.
\end{equation}
Equivalently, the smoothed loss $\ell_{h, \tau}$ can be written as $\ell_{h, \tau} = \rho_\tau \circ K_h$, where $K_h(u) = (1/h)K(u/h)$ and ``$\circ$" denotes the convolution operator.
 Commonly used kernel functions include (i) uniform kernel,  (ii) Gaussian kernel, (iii) logistic kernel, (iv) Epanechnikov kernel,  and (v) triangular kernel. See Section~3 of \cite{HPTZ2020} for the explicit expressions of $\ell_{h,\tau}$ when the above kernels are used.

Statistical properties of convolution smoothed QR have been studied in the context of linear models under fixed-$p$,  growing-$p$ and high-dimensional settings \citep{FGH2019, HPTZ2020, TWZ2021}.
In the low-dimension setting ``$p\ll n$",   \cite{HPTZ2020} showed that the SQR estimator is (asymptotically) first-order equivalent to the QR estimator. Moreover,   the asymptotic normality of SQR holds under a weaker requirement
on dimensionality than needed for QR. In the high-dimensional regime ``$p \gg n$",   \citet{TWZ2021} proved that the SQR-lasso estimator with a properly chosen bandwidth enjoys the same convergence rate as QR-lasso \citep{BC2011}.   With iteratively reweighted $\ell_1$-regularization,  oracle properties can be achieved by SQR  under a weaker signal strength condition than needed for QR.

Computationally,   \cite{TWZ2021} proposed coordinate descent and ADMM-based algorithms that are tailored to the uniform kernel and Gaussian kernel, respectively.  These algorithms do not exhibit evident advantages over those for penalized QR, and also limit the choice of both kernel and penalty functions.
This motivates us to develop a more efficient and flexible algorithm for penalized SQR, which accommodates general kernel functions and a broader range of convex penalties.

\section{Local Adaptive Majorize-minimization Algorithm for Penalized SQR}
\label{sec:psalg}


In this section, we describe a unified algorithm to solve the optimization problem for penalized SQR,  which has a general form
\begin{equation} \label{opteq:1}
\displaystyle{\minimize_{\bbeta \in \mathbb{R}^{p}}~\frac{1}{n} \sum_{i=1}^{n} \ell_{h,\tau}(y_i-\bx_i^T\bbeta) + P(\bbeta)},
\end{equation}
where $P(\bbeta)$ is a generic convex penalty function and $\ell_{h,\tau}(\cdot)$ is the smoothed check loss given in \eqref{sqrloss}. 
In this paper, we focus on the following four widely used convex penalty functions.
\begin{enumerate}
\item[1)]  Weighted lasso  \citep{Tibs1996}: $P(\bbeta) = \sum_{j=1}^p \lambda_j |\beta_j|$, where $\lambda_j \ge0$ for $j=1,\ldots, p$; 
\item[2)] Elastic net \citep{ZH2005}: $P(\bbeta) = \lambda\alpha  \|\bbeta\|_1 + \lambda(1-\alpha) \|\bbeta\|_2^2$,  where $\lambda >0$ is a sparsity-inducing parameter and $\alpha \in [0,1]$ is a user-specified constant that controls the tradeoff between the $\ell_1$ penalty and the ridge penalty;
\item[3)]  Group lasso \citep{YL2006}: $P(\bbeta)  = \lambda \sum_{g=1}^G w_g \|\bbeta_g\|_2$, where $\bbeta = (\bbeta_1^T,\ldots,\bbeta_G^T)^T$ and $\bbeta_g$ is a sub-vector of $\bbeta$ corresponding to the $g$th group of coefficients, and    $w_g>0$ are predetermined weights; 
 \item[4)] Sparse group lasso \citep{SFHT2013}: $P(\bbeta)  =  \lambda\|\bbeta\|_1 +  \lambda\sum_{g=1}^G  w_g \|\bbeta_g\|_2$. 
 \end{enumerate}

We employ the local adaptive majorize-minimization (LAMM) principle to derive an iterative algorithm for solving~\eqref{opteq:1}.  The LAMM principle is a generalization of the majorize-minimization (MM) algorithm \citep{LHY2000,HL2004} to high dimensions, and has been  applied to penalized least squares,  generalized linear models \citep{FLSF2018} and robust regression \citep{PSZ2021}.
We first provide a brief overview of the LAMM algorithm.

Consider the minimization of a general  smooth function $f(\bbeta)$.  Given an estimate $\hat{\bbeta}^{k-1}$ at the $k$th iteration,   the LAMM algorithm locally majorizes $f(\bbeta)$ by a properly constructed function $g(\bbeta | \hat{\bbeta}^{k-1})$ that satisfies  the local property
\begin{equation}
\label{Eq:LAMMPrinciple}
f(\hat{\bbeta}^{k})\le g(\hat{\bbeta}^{k}|\hat{\bbeta}^{k-1})  \qquad \mathrm{and}\qquad g(\hat{\bbeta}^{k-1} | \hat{\bbeta}^{k-1}) = f(\hat{\bbeta}^{k-1}),
\end{equation}
where $\hat{\bbeta}^k = \argmin_{\bbeta}  \,g(\bbeta | \hat{\bbeta}^{k-1}) $.
The ensures the decrease of the objective function after each step,   i.e., $f(\hat{\bbeta}^{k}) \le f(\hat{\bbeta}^{k-1})$.
Note that~\eqref{Eq:LAMMPrinciple} is a relaxation of the global majorization requirement, $f({\bbeta})\le g({\bbeta}|\hat{\bbeta}^{k-1})$, used in the MM algorithm \citep{LHY2000,HL2004}.

Motivated by the local property in~\eqref{Eq:LAMMPrinciple}, we now derive an iterative algorithm for solving~\eqref{opteq:1}.
For notational convenience, let $Q(\bbeta) = n^{-1}\sum_{i=1}^n\ell_{h,\tau}( y_i-\bx_i^T\bbeta)$  and let $\nabla Q(\bbeta)$ be the gradient of $Q(\bbeta)$. 
We locally majorize $Q(\bbeta)$ given $\hat{\bbeta}^{k-1}$ by constructing an isotropic quadratic function of the form
\begin{equation*}
F(\bbeta|\phi_k,\hat{\bbeta}^{k-1}) = Q(\hat{\bbeta}^{k-1}) + \langle \nabla Q(\hat{\bbeta}^{k-1}), \bbeta - \hat{\bbeta}^{k-1} \rangle + \frac{\phi_k}{2} \| \bbeta - \hat{\bbeta}^{k-1} \|_2^2,
\end{equation*}
where $\phi_k > 0$ is a quadratic parameter  (to be determined) at the $k$th iteration. Then, define the $k$th iterate $\hat{\bbeta}^k$ as the solution to
\begin{equation} \label{opteq:2}
\displaystyle{\minimize_{\bbeta \in \mathbb{R}^{p}}~F(\bbeta|\phi_k,\hat{\bbeta}^{k-1}) + P(\bbeta) }.
\end{equation}

To ensure the descent of the objective function in \eqref{opteq:1} at each iteration, the parameter $\phi_k > 0$ needs to be sufficiently large such that $ Q(\hat{\bbeta}^k) \le F(\hat{\bbeta}^k|\phi_k,\hat{\bbeta}^{k-1})$. Consequently,
\begin{equation*}
\begin{split}
Q(\hat{\bbeta}^k) + P(\hat{\bbeta}^k)  & \leq F(\hat{\bbeta}^k|\phi_k,\hat{\bbeta}^{k-1}) + P (\hat{\bbeta}^k)  \\
& \leq F(\hat{\bbeta}^{k-1}|\phi_k,\hat{\bbeta}^{k-1}) + P (\hat{\bbeta}^{k-1}) \\
& = Q(\hat{\bbeta}^{k-1}) +  P (\hat{\bbeta}^{k-1}),
\end{split}
\end{equation*}
where the second inequality is due to the fact that $\hat{\bbeta}^{k}$ is a minimizer of~\eqref{opteq:2}.
In practice, we choose $\phi_k$ by starting from a small value $\phi_0 = 0.01$ and successively inflate it by a factor $\gamma = 1.2$ until the majorization requirement $ Q(\hat{\bbeta}^k) \le F(\hat{\bbeta}^k|\phi_k,\hat{\bbeta}^{k-1}) $ is met at each iteration of the LAMM algorithm.

One of the main advantages of our approach is that the isotropic form of $F(\bbeta|\phi_k,\hat{\bbeta}^{k-1})$, as a function of $\bbeta$, permits a simple analytic solution $\hat{\bbeta}^k = (\hat{\beta}_1^k,  \ldots, \hat{\beta}_p^k)^T$ for different convex penalty functions $P(\bbeta)$.  
By the first-order optimization condition,  $\hat{\bbeta}^k$ satisfies
\$
 \mathbf{0} \in \nabla_{\bbeta} Q(\hat{\bbeta}^{k-1}) + \phi_k (\hat{\bbeta}^k - \hat{\bbeta}^{k-1}) + \partial P(\bbeta) |_{\bbeta = \hat \bbeta^k},
\$
where $\partial P$ denotes the subdifferential of $P: \RR^p\to [0, \infty)$. With certain convex penalties,  a closed-form expression for $\hat{\bbeta}^k$ can be derived from the above condition. Since a common practice is to leave the intercept term unpenalized,  its update  takes a simple form $ \hat{\beta}_1^k = \hat{\beta}_1^{k-1} - \phi_k^{-1} \partial_{\beta_1} Q(\hat{\bbeta}^{k-1})$. 
Detailed update rules of $\hat{\bbeta}^k$ for the above four convex penalties are summarized in  Algorithm~\ref{Alg:general},  in which  $S(a,b) = \sign(a) \cdot (| a| -b)_{+}$ denotes the shrinkage operator, where $\text{sign}(\cdot)$ is the sign function and $(c)_{+} = \text{max}(c, 0)$.  For all the four penalty functions,  the dominant computational effort of each LAMM update is a relatively cheap matrix-vector multiplication involving $\mathbf{X} = (\bx_1, \ldots, \bx_n)^T$,  thus with a complexity $O(np)$.

\begin{algorithm}[!htp]
\small
\caption{ The LAMM Algorithm for Solving (\ref{opteq:1}).}
\label{Alg:general}
\textbf{Input:} kernel function $K(\cdot)$, penalty function $P(\cdot)$, regularization parameters,   bandwidth $h$, inflation factor $\gamma = 1.2$, and convergence criterion $\epsilon$.\\
\textbf{Initialization:}  $\hat{\bbeta}^{0}=\boldsymbol{0}$, $\phi_0 = 0.01$. \\
Repeat the following steps until the stopping criterion $\|\hat{\bbeta}^{k}-\hat{\bbeta}^{k-1}\|_2 \le \epsilon$ is met, where $\hat{\bbeta}^{k}$ is the $k$th iterate.  
\begin{enumerate}
\item Set $\phi_k$ $\leftarrow$ max\{$\phi_0$, $\phi_{k-1}/\gamma$\}.
\item \textbf{repeat}
\item for $j = 1, \ldots, p$ (or $g=1,\ldots,G$ for group lasso and sparse group lasso), update $\hat{\beta}_j^k$ (or $\hat{\bbeta}_g^k$) as follows: 

\begin{tabular}{c|c}
\hline
weighted lasso & $\hat{\beta}_j^k = S(\hat{\beta}_j^{k-1} - \phi_k^{-1} \nabla_{\beta_j} Q(\hat{\bbeta}^{k-1}), \phi_k^{-1} \lambda_j)$.\\
\hline
elastic net & $\hat{\beta}_j^k = \frac{1}{1 + 2\phi_k^{-1}\lambda(1-\alpha)} S(\hat{\beta}_j^{k-1} - \phi_k^{-1} \nabla_{\beta_j} Q(\hat{\bbeta}^{k-1}), \phi_k^{-1} \lambda \alpha)$.\\
\hline
group lasso & $\hat{\bbeta}_g^k = (\hat{\bbeta}_g^{k-1} - \phi_k^{-1} \nabla_{\bbeta_g} Q(\hat{\bbeta}^{k-1}))(1-\frac{\lambda w_g}{\phi_k \| \hat{\bbeta}_g^{k-1} - \phi_k^{-1} \nabla_{\bbeta_g} Q(\hat{\bbeta}^{k-1}) \|_2})_{+}$.\\
\hline
sparse group lasso & $\hat{\bbeta}_g^k = S(\hat{\bbeta}_g^{k-1} - \phi_k^{-1} \nabla_{\bbeta_g} Q(\hat{\bbeta}^{k-1}), \phi_k^{-1} \lambda)(1-\frac{\lambda w_g}{\phi_k \| \hat{\bbeta}_g^{k-1} - \phi_k^{-1} \nabla_{\bbeta_g} Q(\hat{\bbeta}^{k-1}) \|_2})_{+}$.\\
\hline
\end{tabular}
\item \quad \textbf{If} $F(\hat{\bbeta}^k|\phi_k, \hat{\bbeta}^{k-1}) < Q(\hat{\bbeta}^k)$, set $\phi_k = \gamma\phi_k$.
\item \textbf{until} $F(\hat{\bbeta}^k| \phi_k, \hat{\bbeta}^{k-1}) \geq Q(\hat{\bbeta}^k)$.
\end{enumerate}
\textbf{Output:} the updated parameter $\hat{\bbeta}^{k}$.
\end{algorithm}

\section{Numerical Studies}
\label{sec:ns}

In this section,  we perform extensive numerical studies to evaluate the performance of the LAMM algorithm (Algorithm~\ref{Alg:general}) for fitting penalized SQR with four convex penalties,  the lasso ($\ell_1$ penalty), elastic net, group lasso, and sparse group lasso.  
We implement Algorithm~\ref{Alg:general} using the Gaussian kernel.  The numerical performance under other commonly used kernels, such the logistic kernel, Laplacian kernel and uniform kernel, are quite similar and thus  we omit the correspondent results.
As suggested in \citet{TWZ2021}, we set the default bandwidth value as $h=  \text{max}\{0.05, \sqrt{\tau(1-\tau)}(\log p/n)^{1/4}\}$ throughout the numerical studies.  The empirical evidence from \citet{HPTZ2020} and \citet{TWZ2021} shows that the SQR estimator is not susceptible to the choice of $h$ in a reasonable range that is neither too small nor too large.
In Section~\ref{subsec:sd}, we fit penalized SQR with the $\ell_1$ and elastic net penalties on simulated data with sparse regression coefficients.  We also evaluate the computational efficiency of Algorithm~\ref{Alg:general},  implemented by the \texttt{conquer} package,  by comparing it to several state-of-the-art packages on penalized regression.  In Section~\ref{subsec:sd:group}, we fit penalized SQR with the group lasso penalty on simulated data for which the  groups of regression coefficients are sparse.

\subsection{Simulated Data with Sparse Regression Coefficients}
\label{subsec:sd}
We start with generating $\tilde{\bx}_i\in \mathbb{R}^{p}$ from a multivariate normal distribution $N_{p}(\mathbf{0},\bSigma)$, where $\bSigma = (0.7^{|j-k|})_{1 \leq j,k \leq p}$, and set $\bx_i = (1,\tilde{\bx}^T_i)^T$.  
Given $\tau \in (0,1)$, we generate the response $y_i$ from the following linear heteroscedastic model:
\begin{equation}
\label{eq:simmodel}
y_i = \bx_i^T \bbeta^* + (0.5x_{i, p+1} + 1)\{\epsilon_i - F_{\epsilon_i}^{-1}(\tau)\},
\end{equation}
where $F_{\epsilon_i}^{-1}(\tau)$ denotes the $\tau$th quantile of the noise variable $\epsilon_i$.
We consider two noise distributions: (i) Normal distribution with mean zero and variance 2---$\epsilon_i \sim N(0,2)$, and (ii) $t$-distribution with 1.5 degrees of freedom---$\epsilon_i \sim t_{1.5}$.
Moreover, the vector of regression coefficients $\bbeta^*$ takes the following two forms: (i) sparse $\bbeta^*$ with $\beta_1^* = 4$ (intercept), $\beta^*_2=1.8$, $\beta^*_4=1.6$, $\beta^*_6=1.4$, $\beta^*_8=1.2$, $\beta^*_{10} = 1$, $\beta^*_{12} = -1$, $\beta^*_{14} = -1.2$, $\beta^*_{16} = -1.4$, $\beta^*_{18} = -1.6$, $\beta^*_{20} = -1.8$,  and $\beta^*_j=0$ for all other $j$'s, and (ii) dense $\bbeta^*$ with $\beta_1^* = 4$ (intercept), $\beta_j^* = 0.8$ for $j=2,\ldots,100$, and $0$ otherwise.

We compare the proposed algorithm for penalized SQR with the $\ell_1$ and elastic net penalties to the $\ell_1$-penalized QR implemented by the 
\texttt{R} package \texttt{rqPen}{\footnote{The \texttt{rqPen} package does not have the elastic net penalty option.}} \citep{SM2020}.  Note that the ADMM-based algorithm proposed by \cite{Gu2018}, implemented in the \texttt{R} package \texttt{FHDQR}, is incompatible with the current  version of \texttt{R} and hence is not included in this paper.  The regularization parameter $\lambda$ is selected via ten-fold cross-validation for which the validation error is defined through the quantile loss.
We set the additional tuning parameter $\alpha$ for the elastic net penalty as $\alpha = \{0.3,0.5,0.7\}$.
To evaluate the statistical performance of different methods, we report the estimation error under the $\ell_2$-norm, i.e., $\| \hat{\bbeta} - \bbeta^*\|_2$, as well as the true positive rate (TPR) and false positive rate (FPR), which are defined as the proportion of correctly estimated non-zeros and the proportion of falsely estimated non-zeros, respectively. The results for the sparse and dense $\bbeta^*$, averaged over 100 replications, are reported in Tables~\ref{tab:a} and~\ref{tab:b}, respectively.

From Table~\ref{tab:a} we see that when true signals are genuinely sparse, both $\ell_1$-penalized SQR and QR outperform the elastic net SQR (with different $\alpha$ values) in all three facets. Due to the large number of zeros in $\bbeta^*$, the performance of the elastic net SQR deteriorates as $\alpha$ decreases, where $\alpha\in [0,1]$ is a user-specified parameter that balances the $\ell_1$ penalty and the ridge ($\ell_2$) penalty.  
Table~\ref{tab:b} summarizes the results under a dense $\bbeta^*$ that contains 100 non-zeros coordinates.
In this case, the elastic net SQR estimators tend to have lower estimation error and high true positive rate,
suggesting that the elastic net penalty may be beneficial when the signals are dense and the signal-to-noise ratio is relatively low. 


\begin{table}[!t]
	\fontsize{8}{9}\selectfont
	\centering
	\caption{Numerical comparisons under the linear heteroscedastic model~\eqref{eq:simmodel} with sparse regression coefficients for moderate ($n=500$, $p=250$) and high-dimensional settings ($n=250$, $p=500$). The mean (and standard error) of the estimation error calculated under the $\ell_2$-norm, true and false positive rates (TPR and FPR), averaged over 100 replications, are reported.}
	\label{tab:a}
	\begin{tabular}{| c | l | c c c | c c c|}
	\hline
	\multicolumn{8}{|c|}{Linear Heteroscedastic Model with Sparse $\bbeta^*$ and $\tau = 0.5$}\\ \hline
	 & & \multicolumn{3}{c|}{ $(n = 500, p = 250)$} & \multicolumn{3}{c}{ $(n = 250, p = 500)$} \vline\\ \hline
	Noise & Methods & Error & TPR & FPR & Error & TPR & FPR \\
		\hline
&\texttt{SQR} (lasso)&0.507 (0.009)&1 (0)&0.112 (0.006)&0.861 (0.019)&1 (0)&0.070 (0.003)\\
 &\texttt{SQR} (elastic net,  $\alpha=0.7$)&0.943 (0.013)&1 (0)&0.280 (0.006)&1.617 (0.020)&1 (0)&0.161 (0.004)\\
	 &\texttt{SQR} (elastic net,  $\alpha=0.5$)&1.268 (0.014)&1 (0)&0.429 (0.008)&2.119 (0.017)&1 (0)&0.270 (0.006)\\
$N(0,2)$	 &\texttt{SQR} (elastic net,  $\alpha=0.3$)&1.602 (0.013)&1 (0)&0.635 (0.007)&2.586 (0.013)&1 (0)&0.452 (0.013)\\
	 &\texttt{rqPen} (lasso)&0.533 (0.010)&1 (0)&0.117 (0.006)&0.915 (0.020)&1 (0)&0.076 (0.004)\\

	\hline
&\texttt{SQR} (lasso)&0.454 (0.010)&1 (0)&0.092 (0.004)&0.825 (0.019)&1 (0)&0.065 (0.003)\\
	 &\texttt{SQR} (elastic net,  $\alpha=0.7$)&0.899 (0.015)&1 (0)&0.265 (0.005)&1.695 (0.020)&1 (0)&0.151 (0.004)\\
	 &\texttt{SQR} (elastic net,  $\alpha=0.5$)&1.263 (0.016)&1 (0)&0.419 (0.008)&2.265 (0.017)&1 (0)&0.231 (0.005)\\
	 $t_{1.5}$ 	 &\texttt{SQR} (elastic net,  $\alpha=0.3$)&1.642 (0.016)&1 (0)&0.620 (0.008)&2.725 (0.013)&1 (0)&0.397 (0.006)\\
	 &\texttt{rqPen} (lasso)&0.468 (0.011)&1 (0)&0.112 (0.005)&0.851 (0.020)&1 (0)&0.070 (0.003)\\
	 \hline \hline 
	\multicolumn{8}{|c|}{Linear Heteroscedastic Model with Sparse $\bbeta^*$ and $\tau = 0.7$}\\ \hline	 
	
&\texttt{SQR} (lasso)&0.530 (0.011)&1 (0)&0.110 (0.005)&0.895 (0.020)&1 (0)&0.066 (0.003)\\
	 &\texttt{SQR} (elastic net,  $\alpha=0.7$)&0.988 (0.014)&1 (0)&0.269 (0.006)&1.676 (0.020)&1 (0)&0.159 (0.004)\\
	 &\texttt{SQR} (elastic net,  $\alpha=0.5$)&1.321 (0.016)&1 (0)&0.420 (0.008)&2.189 (0.018)&1 (0)&0.250 (0.005)\\
	$N(0,2)$	 &\texttt{SQR} (elastic net,  $\alpha=0.3$)&1.672 (0.015)&1 (0)&0.615 (0.008)&2.647 (0.013)&1 (0)&0.407 (0.010)\\
	 &\texttt{rqPen} (lasso)&0.553 (0.012)&1 (0)&0.120 (0.005)&0.947 (0.021)&1 (0)&0.070 (0.003)\\

	\hline
 &\texttt{SQR} (lasso)&0.561 (0.014)&1 (0)&0.102 (0.004)&1.023 (0.024)&0.999 (0.001)&0.063 (0.003)\\
	 &\texttt{SQR} (elastic net,  $\alpha=0.7$)&1.073 (0.017)&1 (0)&0.258 (0.006)&1.873 (0.021)&0.999 (0.001)&0.153 (0.004)\\
	 &\texttt{SQR} (elastic net,  $\alpha=0.5$)&1.454 (0.018)&1 (0)&0.398 (0.008)&2.421 (0.018)&1 (0)&0.225 (0.004)\\
	 $t_{1.5}$ 	 &\texttt{SQR} (elastic net,  $\alpha=0.3$)&1.842 (0.017)&1 (0)&0.595 (0.008)&2.863 (0.014)&1 (0)&0.369 (0.005)\\
	 &\texttt{rqPen} (lasso)&0.582 (0.015)&1 (0)&0.103 (0.005)&1.060 (0.026)&0.998 (0.001)&0.068 (0.003)\\

	\hline
	\end{tabular}
\end{table}

\begin{table}[!htp]
	\fontsize{8}{9}\selectfont
	\centering
	\caption{ Numerical comparisons under the linear heteroscedastic model~\eqref{eq:simmodel} with dense regression coefficients.  Other details are as in Table~\ref{tab:a}.}
	\label{tab:b}
	\begin{tabular}{| c | l | c c c | c c c|}
	\hline
	\multicolumn{8}{|c|}{Linear Heteroscedastic Model with Dense $\bbeta^*$ and $\tau = 0.5$}\\ \hline
	 & & \multicolumn{3}{c|}{ $(n = 500, p = 250)$} & \multicolumn{3}{c}{ $(n = 250, p = 500)$} \vline\\ \hline
	Noise & Methods & Error & TPR & FPR & Error & TPR & FPR \\
		\hline
&\texttt{SQR} (lasso)&1.441 (0.015)&1 (0)&0.246 (0.009)&3.018 (0.034)&0.996 (0.001)&0.218 (0.011)\\
 &\texttt{SQR} (elastic net,  $\alpha=0.7$)&1.243 (0.013)&1 (0)&0.322 (0.009)&2.184 (0.028)&1 (0)&0.259 (0.006)\\
	 &\texttt{SQR} (elastic net,  $\alpha=0.5$)&1.210 (0.014)&1 (0)&0.440 (0.009)&2.192 (0.034)&1 (0)&0.407 (0.008)\\
$N(0,2)$	 &\texttt{SQR} (elastic net,  $\alpha=0.3$)&1.260 (0.014)&1 (0)&0.633 (0.007)&2.787 (0.037)&1 (0)&0.718 (0.009)\\
	 &\texttt{rqPen} (lasso)&1.512 (0.015)&1 (0)&0.246 (0.010)&3.239 (0.039)&0.993 (0.001)&0.175 (0.006)\\
\hline
&\texttt{SQR} (lasso)&1.575 (0.020)&1 (0)&0.216 (0.008)&4.340 (0.188)&0.976 (0.002)&0.163 (0.006)\\
	 &\texttt{SQR} (elastic net,  $\alpha=0.7$)&1.296 (0.015)&1 (0)&0.303 (0.008)&2.479 (0.043)&1 (0)&0.239 (0.005)\\
	 &\texttt{SQR} (elastic net,  $\alpha=0.5$)&1.247 (0.014)&1 (0)&0.434 (0.007)&2.406 (0.029)&1 (0)&0.373 (0.005)\\
	 $t_{1.5}$ 	 &\texttt{SQR} (elastic net,  $\alpha=0.3$)&1.316 (0.014)&1 (0)&0.636 (0.006)&2.709 (0.029)&1 (0)&0.594 (0.005)\\
	 &\texttt{rqPen} (lasso)&1.628 (0.022)&1 (0)&0.234 (0.009)&4.012 (0.055)&0.970 (0.002)&0.142 (0.004)\\
\hline \hline 
	\multicolumn{8}{|c|}{Linear Heteroscedastic Model with Dense $\bbeta^*$ and $\tau = 0.7$}\\ \hline	 
	
&\texttt{SQR} (lasso)&1.508 (0.015)&1 (0)&0.223 (0.008)&3.054 (0.034)&0.995 (0.001)&0.194 (0.008)\\
	 &\texttt{SQR} (elastic net,  $\alpha=0.7$)&1.289 (0.012)&1 (0)&0.309 (0.008)&2.222 (0.024)&1 (0)&0.248 (0.005)\\
	 &\texttt{SQR} (elastic net,  $\alpha=0.5$)&1.237 (0.012)&1 (0)&0.431 (0.007)&2.174 (0.025)&1 (0)&0.383 (0.006)\\
	$N(0,2)$	 &\texttt{SQR} (elastic net,  $\alpha=0.3$)&1.291 (0.012)&1 (0)&0.634 (0.006)&2.782 (0.036)&1 (0)&0.711 (0.010)\\
	 &\texttt{rqPen} (lasso)&1.586 (0.016)&1 (0)&0.228 (0.009)&3.356 (0.039)&0.989 (0.001)&0.166 (0.005)\\
	\hline
 &\texttt{SQR} (lasso)&1.912 (0.029)&1 (0)&0.219 (0.007)&4.896 (0.208)&0.961 (0.003)&0.172 (0.007)\\
	 &\texttt{SQR} (elastic net,  $\alpha=0.7$)&1.487 (0.015)&1 (0)&0.302 (0.007)&2.743 (0.043)&0.999 (0)&0.245 (0.005)\\
	 &\texttt{SQR} (elastic net,  $\alpha=0.5$)&1.410 (0.014)&1 (0)&0.431 (0.007)&2.600 (0.033)&1 (0)&0.372 (0.005)\\
	 $t_{1.5}$ 	 &\texttt{SQR} (elastic net,  $\alpha=0.3$)&1.491 (0.016)&1 (0)&0.643 (0.006)&2.889 (0.036)&1 (0)&0.597 (0.006)\\
	 &\texttt{rqPen} (lasso)&1.947 (0.025)&1 (0)&0.226 (0.008)&4.543 (0.058)&0.942 (0.004)&0.129 (0.004)\\
	\hline
	\end{tabular}
\end{table}

\begin{figure}[!t]
\centering
\subfigure[Estimation error for model~\eqref{eq:simmodel} with $N(0,2)$ error.]{\label{fig:esterrorgaussian}\includegraphics[width=.47\linewidth]{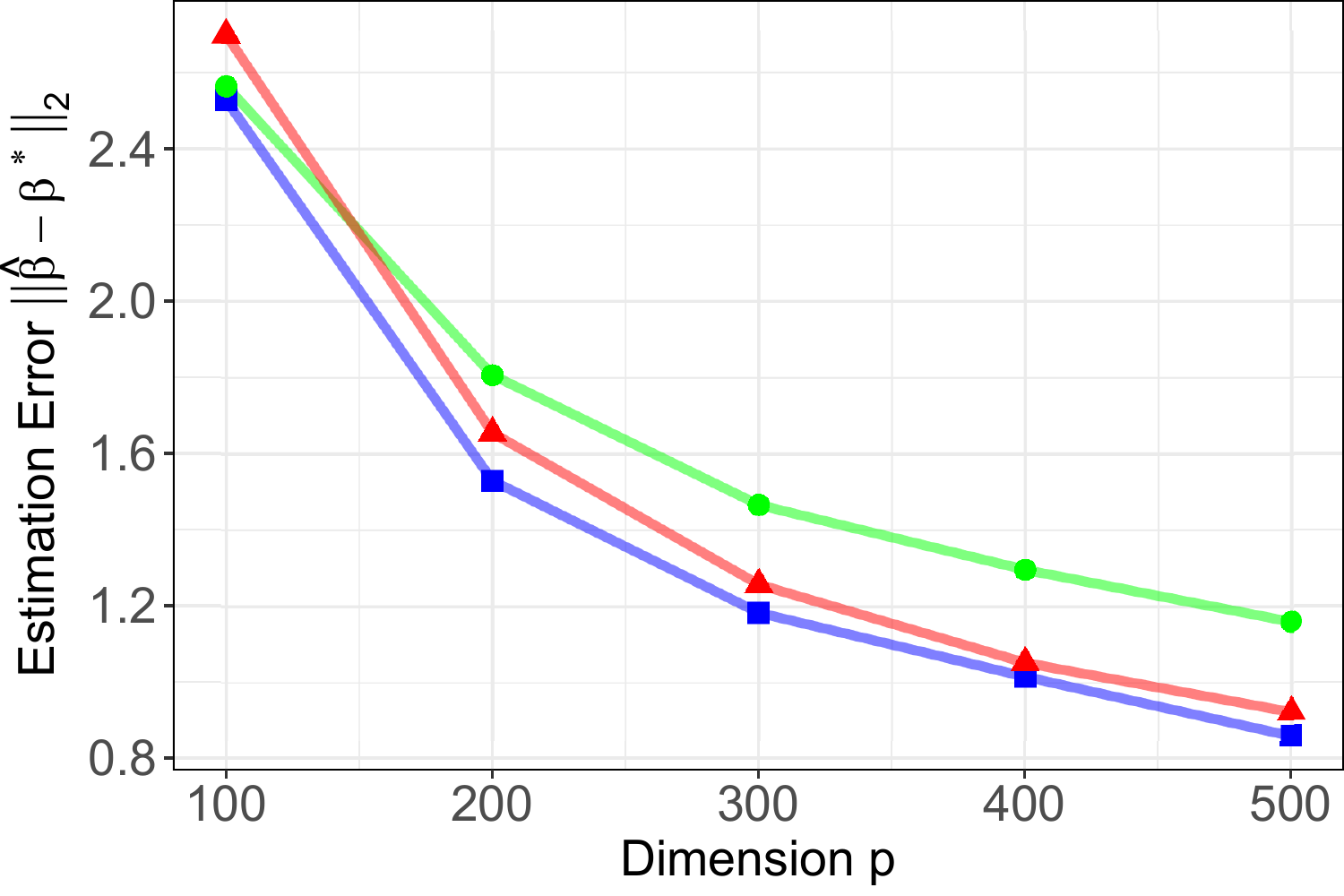}}\qquad
\subfigure[Elapsed time for model~\eqref{eq:simmodel} with $N(0,2)$ error.]{\label{fig:timecompgaussian}\includegraphics[width=.47\linewidth]{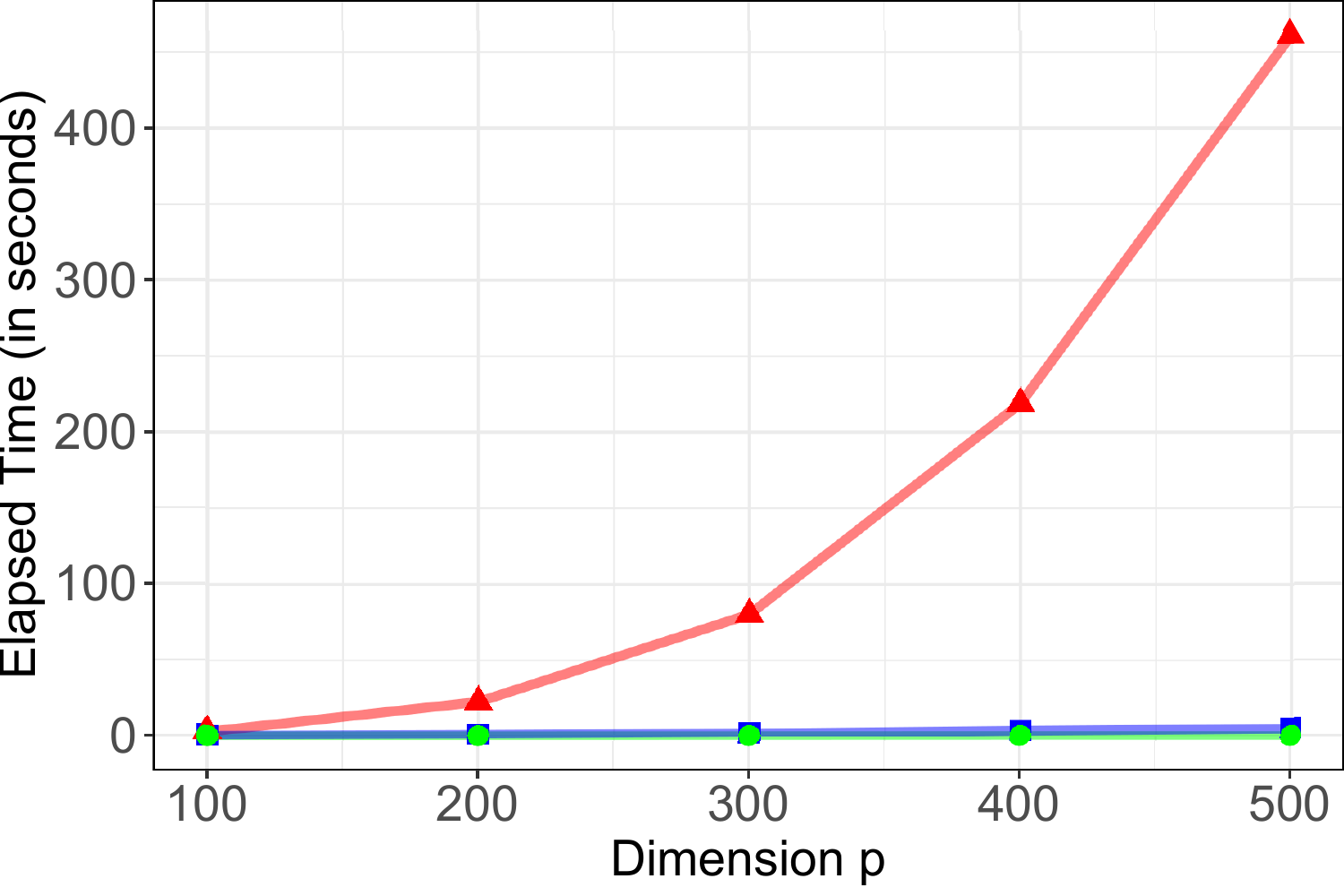}}\\
\subfigure[Estimation error for model~\eqref{eq:simmodel} with $t_{1.5}$ error.]{\label{fig:esterrort}\includegraphics[width=.47\linewidth]{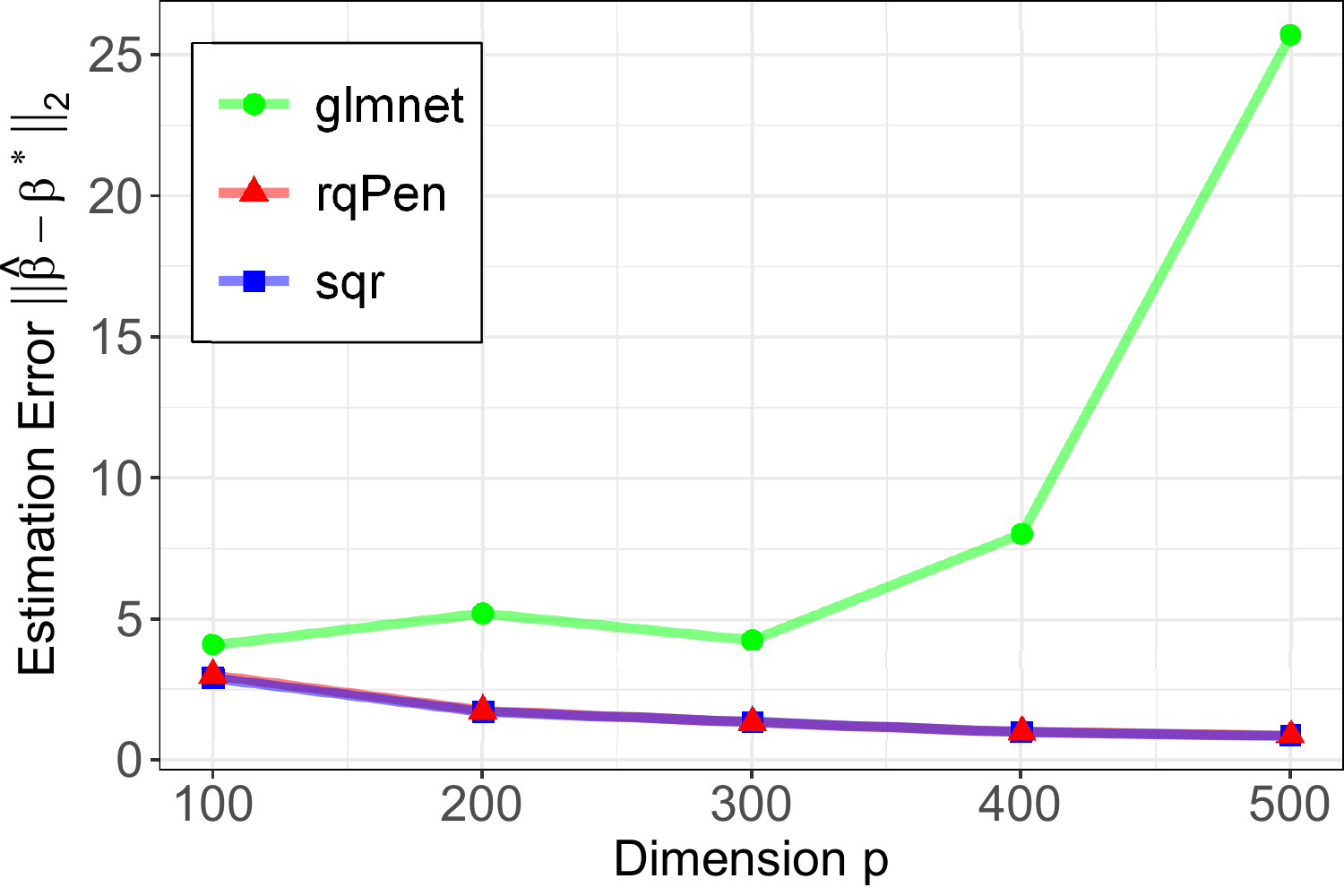}}\qquad
\subfigure[Elapsed time for model~\eqref{eq:simmodel} with $t_{1.5}$ error.]{\label{fig:timecompt}\includegraphics[width=.47\linewidth]{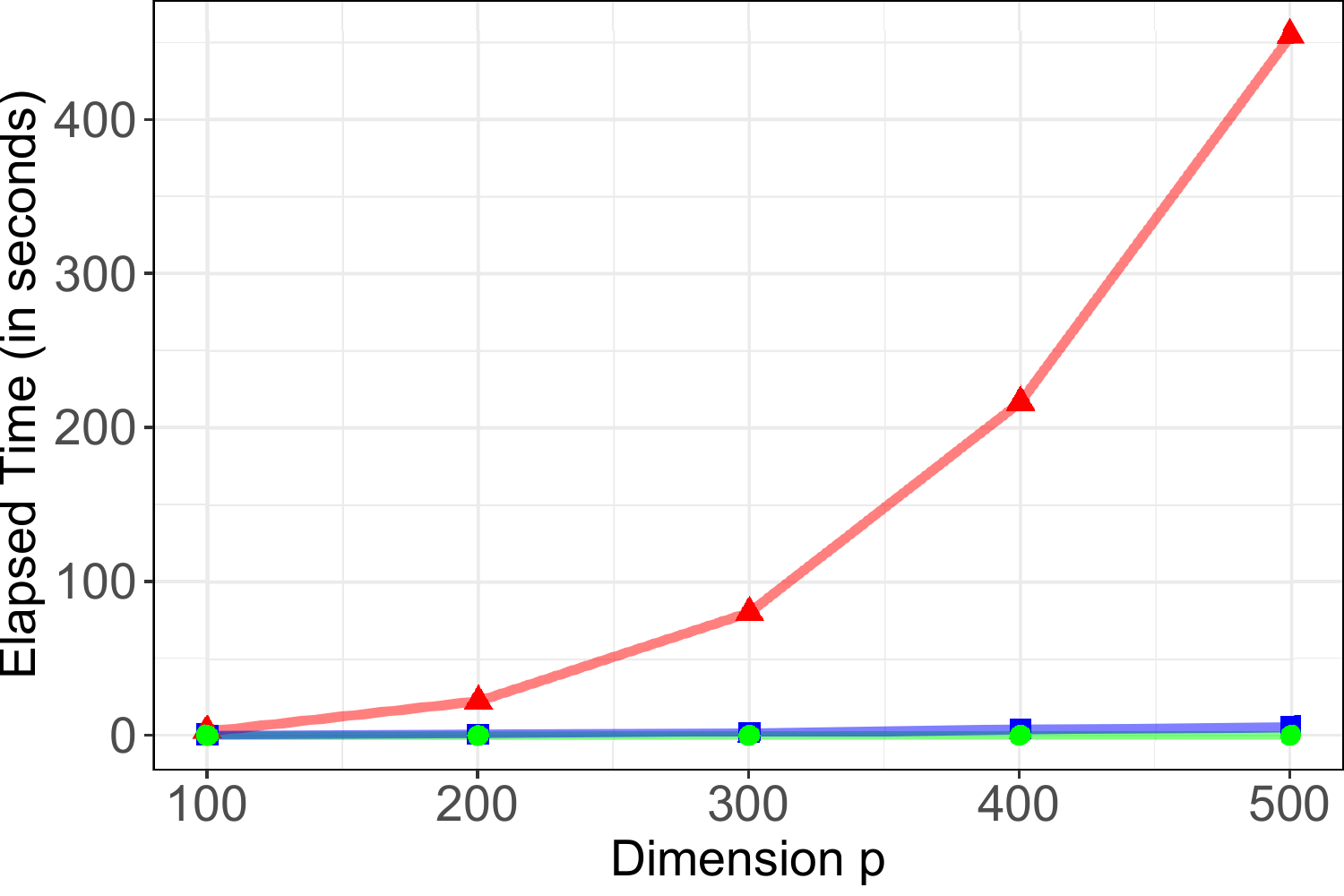}}%
\caption{Estimation error and elapsed time in seconds under model~\eqref{eq:simmodel} with $N(0,2)$ and $t_{1.5}$ random noise and $\tau=0.5$, averaged over 100 data sets for three different algorithms: (i) the propose algorithm for fitting $\ell_1$-penalized SQR with Gaussian kernel; (ii) the $\ell_1$-penalized QR implemented using \texttt{rqPen}; and (iii) the $\ell_1$-penalized least squares method implemented using \texttt{glmnet}.}
\label{fig:comparisons}
\end{figure}

Furthermore, we provide speed comparisons of the three \texttt{R} packages, \texttt{glmnet}, \texttt{rqPen} and \texttt{conquer}, for fitting sparse linear models. As a benchmark, \texttt{glmnet} is used to compute the $\ell_1$-penalized least squares (lasso) estimator, whereas \texttt{rqPen} and \texttt{conquer} are used to fit penalized quantile regression with $\tau=0.5$. For each method, the regularization parameter is selected from a sequence of 50 $\lambda$-values via ten-fold cross-validation.  The curves in panels (a) and (c) of Figure~\ref{fig:comparisons} represent the estimation error (under $\ell_2$ norm) as a function of dimension $p$, and the curves in panels (b) and (d) of Figure~\ref{fig:comparisons} represent the computational time (in second) as a function of dimension $p$. The sample size $n$ is taken to be $2p$.
Under normal errors, the three estimators, lasso, QR-lasso and SQR-lasso, converge at similar rates as $n$ grows; while the lasso estimator is inconsistent when the error follows the $t_{1.5}$-distribution. The SQR-lasso may even slightly outperforms the QR-lasso, indicating that smoothing can potentially improve finite-sample performance. In terms of computational efficiency, our \texttt{conquer} package exhibits a significant improvement over \texttt{rqPen} especially when $p$ is large, and is almost as efficient as \texttt{glmnet}.

\subsection{Simulated Data with Sparse Groups of Regression Coefficients}
\label{subsec:sd:group}

We further assess the performance of penalized SQR with the group lasso penalty by comparing \texttt{conquer} with the  \texttt{R} package \texttt{sgl}. The latter is used to compute the group lasso estimator. 
To this date, we are not aware of any existing \texttt{R} package that implements group lasso penalized quantile regression.  The regularization parameter $\lambda$ is once again selected by ten-fold  cross-validation, and the weights $w_1,\ldots,w_G$ are set as $w_g = \sqrt{p_g}$, where $p_g$ is the dimension of the sub-vector $\bbeta_g$.   

We generate the data according to~\eqref{eq:simmodel} with 10 groups of regression coefficients $\bbeta^*$.  Specifically, we construct a block-diagonal covariance matrix $\bSigma' = \mathrm{diag}(\bSigma_1,\ldots,\bSigma_{10})$, where $\bSigma_1,\bSigma_2\in\RR^{5\times5}$, $\bSigma_3,\bSigma_4,\bSigma_5 \in\RR^{10\times10}$, and $\bSigma_6,\ldots,\bSigma_{15}\in\RR^{(p-40)/10\times(p-40)/10}$, and each block is an exchangeable covariance matrix with diagonal 1 and off-diagonal elements 0.6.
We then generate the covariates from the multivariate normal distribution, $\tilde{\bx}_i \sim N_p(\mathbf{0},\bSigma')$ and set $\bx_i = (1,\tilde{\bx}_i^T)^T$.
We construct $\bbeta^*$ that has a sparse group structure, i.e., $\bbeta^* =\{\beta_0^*,(\bbeta_1^*)^T,\ldots,(\bbeta_{15}^*)^T\}^T$ with $\beta_0^* = 4$ (intercept), $\bbeta_1^* =  {\bf 2} \in \RR^5$, $\bbeta_2^* ={\bf 1.6}\in \RR^5$, $\bbeta_3^* = {\bf -2}\in \RR^{10}$, $\bbeta_4^* =  {\bf 1} \in \RR^{10}$, $\bbeta_5^* = {\bf 0.6}\in \RR^{10}$, and $\bbeta_6^* =  \cdots = \bbeta_{15}^* = {\bf 0} \in \RR^{10}$.

To assess the performance of group lasso SQR, we calculate the group TPR and group FPR, defined as the proportion of groups that are correctly estimated to contain non-zeros, and the proportion of groups that are incorrectly estimated to contain non-zeros, respectively. Since $\ell_1$-penalized methods do not induce group structures, the group TPR and FPR are not well defined.
Results under Gaussian and $t$-distributed random noise, averaged over 100 replications, are reported in Table~\ref{tab:c}.
Under sparse group structures, the group lasso SQR achieves the best performance cross all settings as expected. This indicates that using the group lasso penalty may be beneficial when the covariates are highly correlated within predefined groups.

\begin{table}[!t]
	\fontsize{8}{9}\selectfont
	\centering
	\caption{ Numerical comparisons under the linear heteroscedastic model~\eqref{eq:simmodel} with groups of regression coefficients for moderate ($n=500$, $p=250$) and high-dimensional settings ($n=250$, $p=500$). The mean (and standard error) of the estimation error calculated under the $\ell_2$-norm, group true and group false positive rates (Group TPR and Group FPR), averaged over 100 replications, are reported.}
	\label{tab:c}
	\begin{tabular}{| c | l | c c c | c c c|}
	\hline
	\multicolumn{8}{|c|}{Linear Heteroscedastic Model with Sparse Groups of  $\bbeta^*$ and $\tau = 0.5$}\\ \hline
	 & & \multicolumn{3}{c|}{ $(n = 500, p = 250)$} & \multicolumn{3}{c}{ $(n = 250, p = 500)$} \vline\\ \hline
	Noise & Methods & Error & Group TPR & Group  FPR & Error & Group  TPR & Group  FPR \\
		\hline
&\texttt{SQR} (lasso)&0.707 (0.009)&NA&NA&1.154 (0.016)&NA & NA\\
	  $N(0,2)$	 &\texttt{rqPen} (lasso)&0.712 (0.009)&NA&NA&1.223 (0.021)&NA & NA\\
         &\texttt{SQR} (group)&0.527 (0.006)&1 (0)&0.018 (0.004)&0.745 (0.010)&1 (0)&0.046 (0.009)\\

	 &\texttt{SGL} (group)&1.009 (0.044)&1 (0)&0.061 (0.011)&1.395 (0.069)&1 (0)&0.070 (0.009)\\

	\hline
&\texttt{SQR} (lasso)&0.948 (0.032)&NA & NA&1.510 (0.042)&NA & NA\\
	 $t_{1.5}$	  &\texttt{rqPen} (lasso)&0.663 (0.013)&NA & NA&1.246 (0.023)&NA & NA\\
         &\texttt{SQR} (group)&0.662 (0.022)&1 (0)&0.001 (0.001)&0.875 (0.020)&1 (0)&0.003 (0.002)\\

	 &\texttt{SGL} (group)&2.502 (0.369)&1 (0)&0.139 (0.027)&2.435 (0.151)&1 (0)&0.082 (0.019)\\ \hline
	\multicolumn{8}{|c|}{Linear Heteroscedastic Model with Sparse Groups of  $\bbeta^*$ and $\tau = 0.7$}\\ \hline	 
	
&\texttt{SQR} (lasso)&0.753 (0.011)&NA & NA&1.252 (0.019)&NA & NA\\
	 $N(0,2)$	  &\texttt{rqPen} (lasso)&0.769 (0.011)&NA & NA&1.299 (0.020)&NA & NA\\
         &\texttt{SQR} (group)&0.550 (0.007)&1 (0)&0.019 (0.005)&0.780 (0.011)&1 (0)&0.050 (0.009)\\

	 &\texttt{SGL} (group)&1.209 (0.053)&1 (0)&0.181 (0.012)&1.585 (0.080)&1 (0)&0.181 (0.015)\\
	\hline
&\texttt{SQR} (lasso)&1.223 (0.044)&NA & NA&1.896 (0.062)&NA & NA\\
	  $t_{1.5}$	 &\texttt{rqPen} (lasso)&0.822 (0.014)&NA & NA&1.523 (0.028)&NA & NA\\
         &\texttt{SQR} (group)&0.862 (0.033)&1 (0)&0.001 (0.001)&1.074 (0.030)&1 (0)&0.004 (0.002)\\
	 &\texttt{SGL} (group)&2.591 (0.369)&1 (0)&0.145 (0.028)&2.537 (0.152)&1 (0)&0.089 (0.019)\\
	\hline
	\end{tabular}
\end{table}

\section{Fused Lasso Additive SQR and World Happiness Data }\label{sec:5}

In this section, we employ the fused lasso additive smoothed quantile regression for flexible and interpretable modeling of the conditional relationship between country happiness level and a set of covariates, using the world happiness data \citep{UNDP2012, WBG2012, HLS2013} previously studied in \cite{PWS2016}. Specifically, the goal is to study the conditional distribution of country-level happiness index, the average of Cantril Scale \citep{C1965} responses of approximately 3000 residents in each country, given twelve country-level predictors, including but are not limited to log gross national income (USD), satisfaction with freedom of choice, satisfaction with job, satisfaction with community, and trust in national government.

We first provide a brief overview of the fused lasso additive model in \cite{PWS2016}.   
The fused lasso additive model seeks to balance interpretability and flexibility by approximating the additive function for each covariate via a piecewise constant function.
Let $\btheta_j = (\theta_{1j},\ldots,\theta_{nj})^T$ for $j=1,\ldots,p$ and let $\Db$ be an $(n-1) \times n$ matrix with entries $D_{ii}=1$ and $D_{i(i+1)}= -1$ for $i=1,\ldots,n-1$, and $D_{ij} = 0$ for $j \ne i,i+1$.
Moreover, let $\Pb_j$ be a permutation matrix that orders the elements of $\bx_j = (x_{1j},\ldots, x_{nj})^T$ from least to greatest.
The fused lasso additive model estimator in \citet{PWS2016} can be obtained by solving the convex optimization problem 
\begin{equation} \label{opteq:flam}
\displaystyle{\minimize_{\theta_0 \in \mathbb{R}, \btheta_j \in \mathbb{R}^n} ~\frac{1}{n} \sum_{i=1}^{n} \Bigg(y_i - \theta_0 - \sum_{j=1}^p \theta_{ij}\Bigg)^2 + \lambda \sum_{j=1}^p \| \Db \Pb_j \btheta_j \|_1}, \quad \text{subject to } \mathbbm{1}^T\btheta_{j} = 0 \;\forall j,
\end{equation}
where $\| \Db \Pb_j \btheta_j \|_1$ is a fused lasso type penalty that encourages the consecutive entries of the ordered parameters $\mathbf{P}_j\btheta_j$ to be the same. 
We refer the reader to \citet{PW2019}, \citet{WW2019}, and \citet{ST2019} for a summary of recent work on flexible and interpretable additive models.

We now propose the fused lasso additive smoothed quantile regression for interpretable and flexible modeling of the conditional distribution of $y$ given $\bx$ at specific quantile levels. 
Specifically, we propose to solve the following optimization problem by substituting  the squared error loss in~\eqref{opteq:flam} via the smoothed quantile loss in~\eqref{sqrloss}:
\begin{equation} \label{opteq:3}
\displaystyle{\minimize_{\theta_0 \in \RR, \btheta_j \in \mathbb{R}^n} ~\frac{1}{n} \sum_{i=1}^{n} \ell_{h,\tau}\left(y_i - \theta_0 - \sum_{j=1}^p \theta_{ij}\right) + \lambda \sum_{j=1}^p \| \Db \Pb_j \btheta_j \|_1}, \quad \text{subject to } \mathbbm{1}^T\btheta_{j} = 0 \;\forall j.
\end{equation}
Optimization problem~\eqref{opteq:3} is convex and can be solved using a block coordinate descent algorithm similar to that of \citet{PWS2016}, which we outline in Algorithm~\ref{Alg:bcd}.
Each iteration in Algorithm~\ref{Alg:bcd} involves solving a penalized smoothed quantile regression problem with a fused lasso type penalty~\eqref{alg.opt.eq}. 
\citet{TT2011} showed that the fused lasso type penalty in~\eqref{alg.opt.eq} can be rewritten as a weighted lasso penalty through some  transformation on the regression coefficients, and thus Algorithm~\ref{Alg:general} can naturally be employed to solve~\eqref{alg.opt.eq}.   We omit the derivations and refer the reader to \citet{TT2011} for the details. 

\begin{algorithm}[!htp]
	\small
	\caption{ A Block Coordinate Descent Algorithm for Solving~\eqref{opteq:3}.}
	\label{Alg:bcd}
	\textbf{Input:} kernel function $K(\cdot)$, regularization parameter $\lambda$, smoothing bandwidth parameter $h$, and convergence criterion $\epsilon$.\\
	\textbf{Initialization:}  $\hat{\theta}_0^{0}=0$, $\hat{\btheta}_j^{0}=\boldsymbol{0}$ \;for $j=1,\ldots,p$.\\
	\textbf{Iterate:} for each $j = 1,\ldots,p$, until the stopping criterion $|\hat{\theta}_0^k - \hat{\theta}_0^{k-1}| + \sum_{j=1}^p \|\hat{\btheta}_j^k-\hat{\btheta}_j^{k-1}\|_2  \le \epsilon$ is met, where $\hat{\btheta}_j^{k}$ is the value of $\btheta_j$ obtained at the $k$th iteration.  
	\begin{enumerate}
		\item Update the residual $\rb_j^{k} = \yb - \hat{\theta}_0^{k-1} - \sum_{j'\neq j} \hat{\btheta}_{j'}^{k-1}$.
		\item Update $\hat{\btheta}_j^k$ as
		\begin{equation}\label{alg.opt.eq}
		\hat{\btheta}_j^{k}  =  \displaystyle{\mathrm{argmin}_{\btheta_j \in \mathbb{R}^n} \frac{1}{n} \sum_{i=1}^{n} \ell_{h,\tau}(\rb_j^{k}- \btheta_j) + \lambda\|\Db\Pb_j\btheta_j\|_1}.
		\end{equation}
		\item Update the intercept $\hat{\theta}_0^{k} = \hat{\theta}_0^{k-1} + \text{mean}(\hat{\btheta}_j^{k})$. 
		\item Center the parameters $\hat{\btheta}_j^{k} = \hat{\btheta}_j^{k} - \text{mean}(\hat{\btheta}_j^{k})$.
	\end{enumerate}
	\textbf{Output:} the updated parameters $\hat{\theta}_0^k, \hat{\btheta}_1^k ,\ldots, \hat{\btheta}_p^{k}$.
\end{algorithm}

\begin{figure}[!htp]
	\centering
	\subfigure{\label{fig:plot1}\includegraphics[width=.264\linewidth]{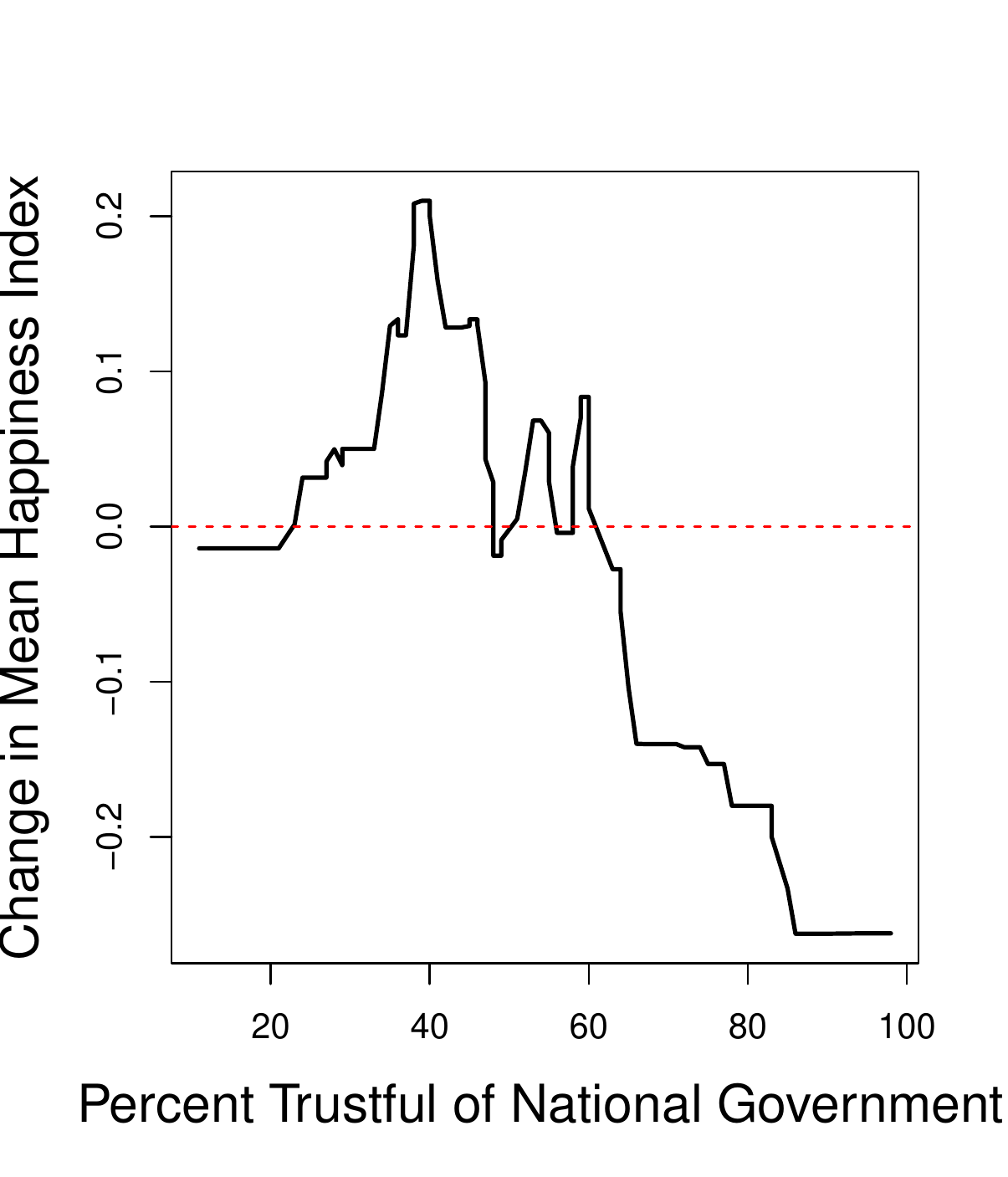}}
	\subfigure{\label{fig:plot2}\includegraphics[width=.264\linewidth]{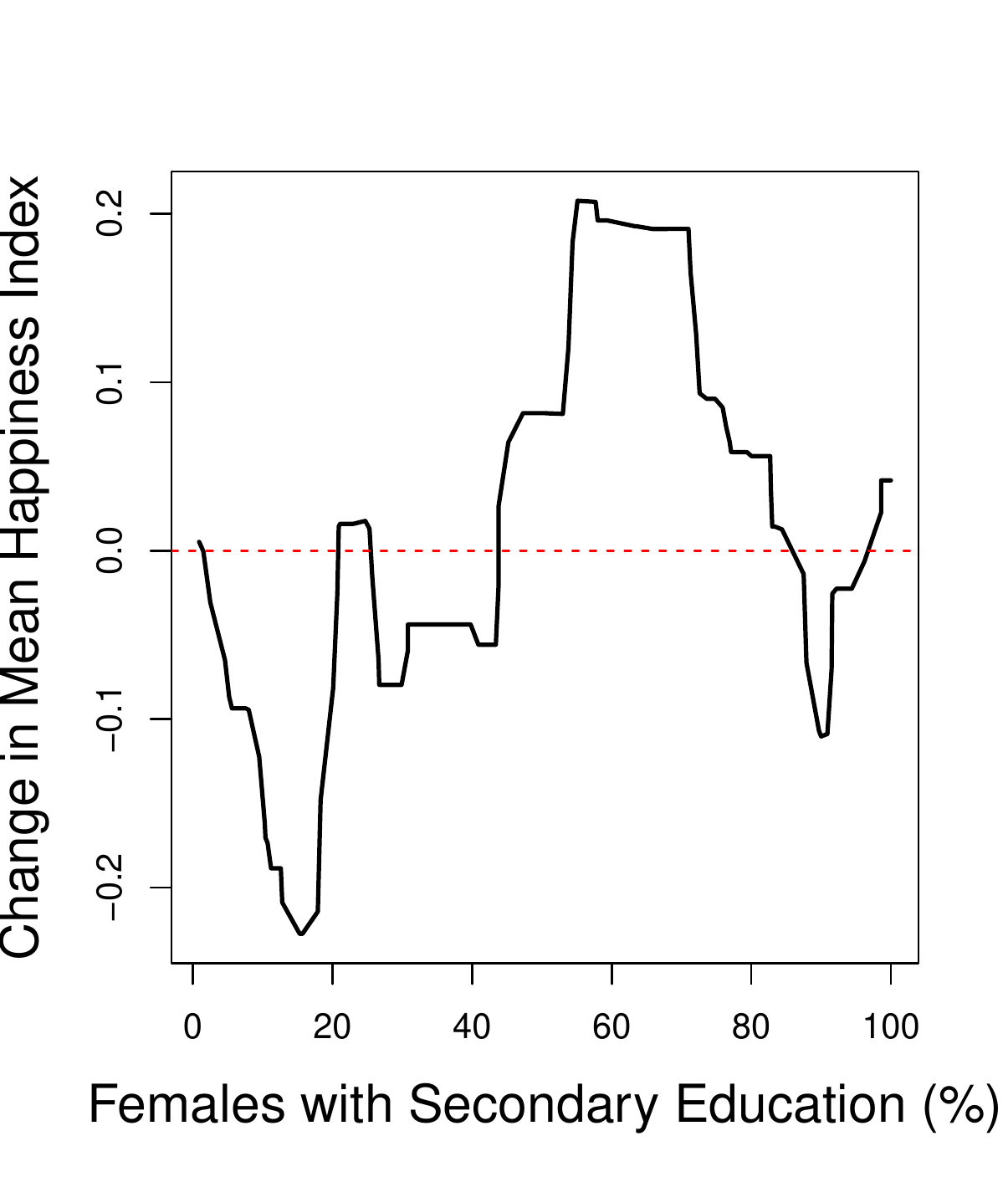}}
	\subfigure{\label{fig:plot3}\includegraphics[width=.264\linewidth]{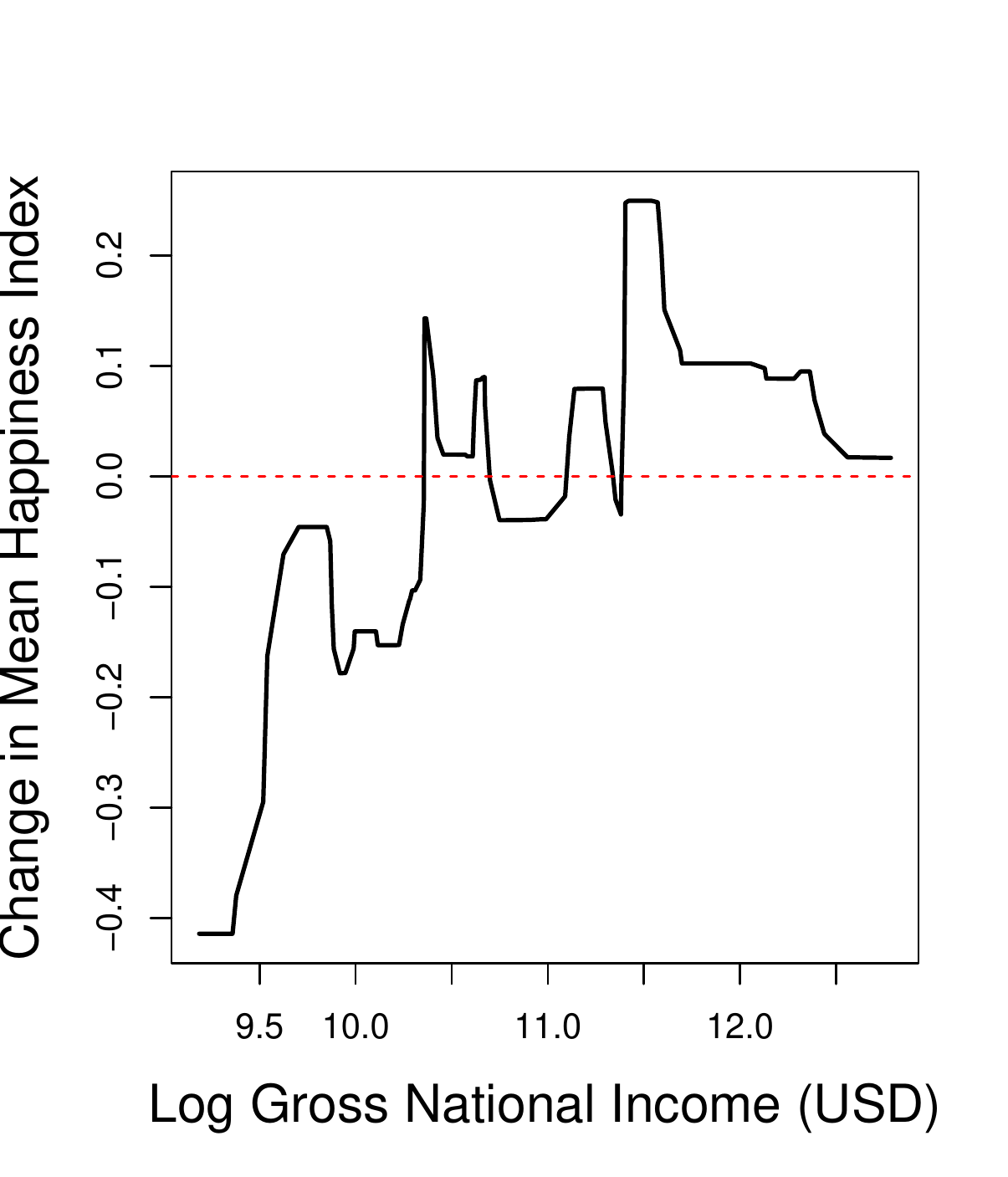}}
	\subfigure{\label{fig:plot5}\includegraphics[width=.264\linewidth]{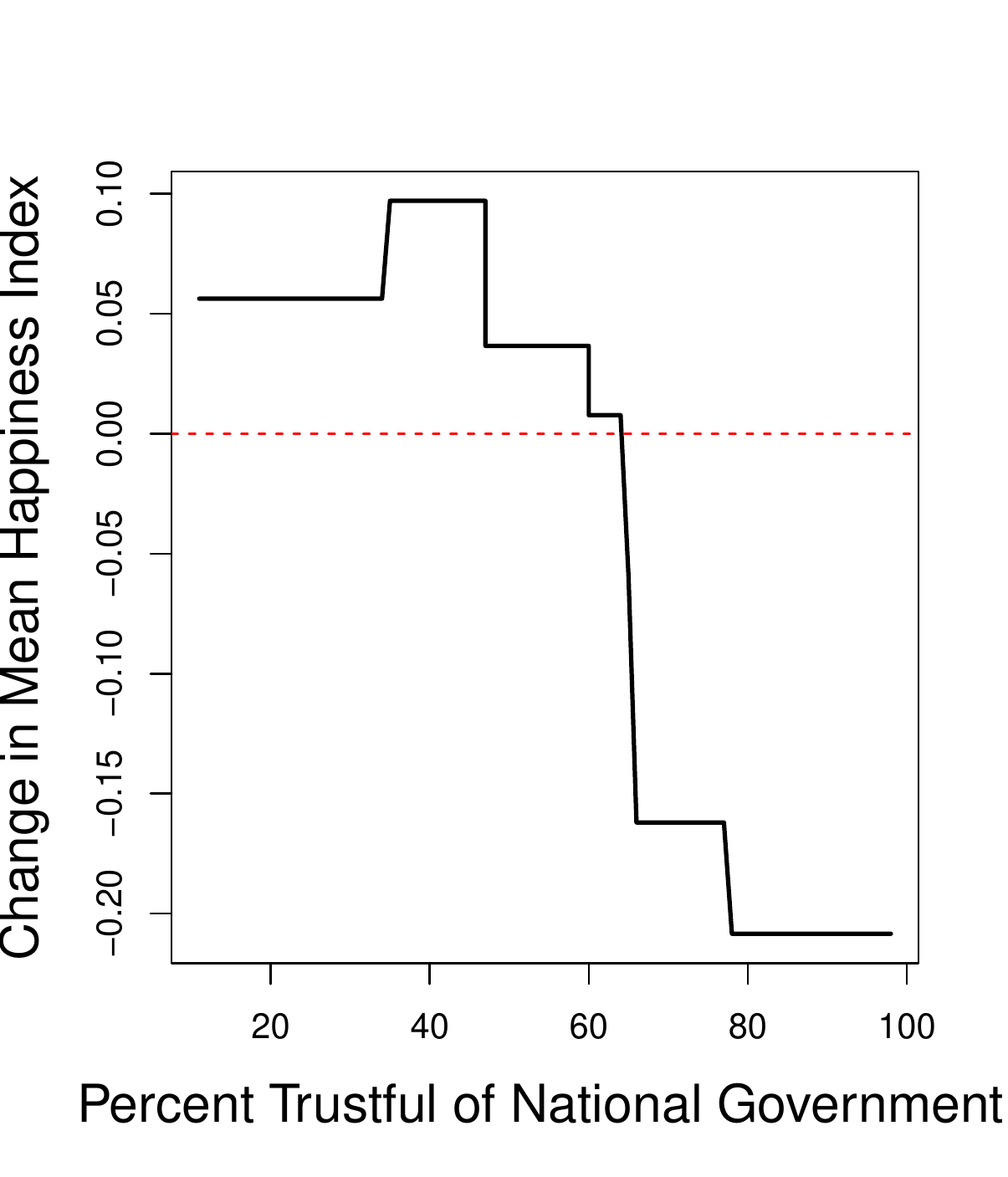}}
	\subfigure{\label{fig:plot6}\includegraphics[width=.264\linewidth]{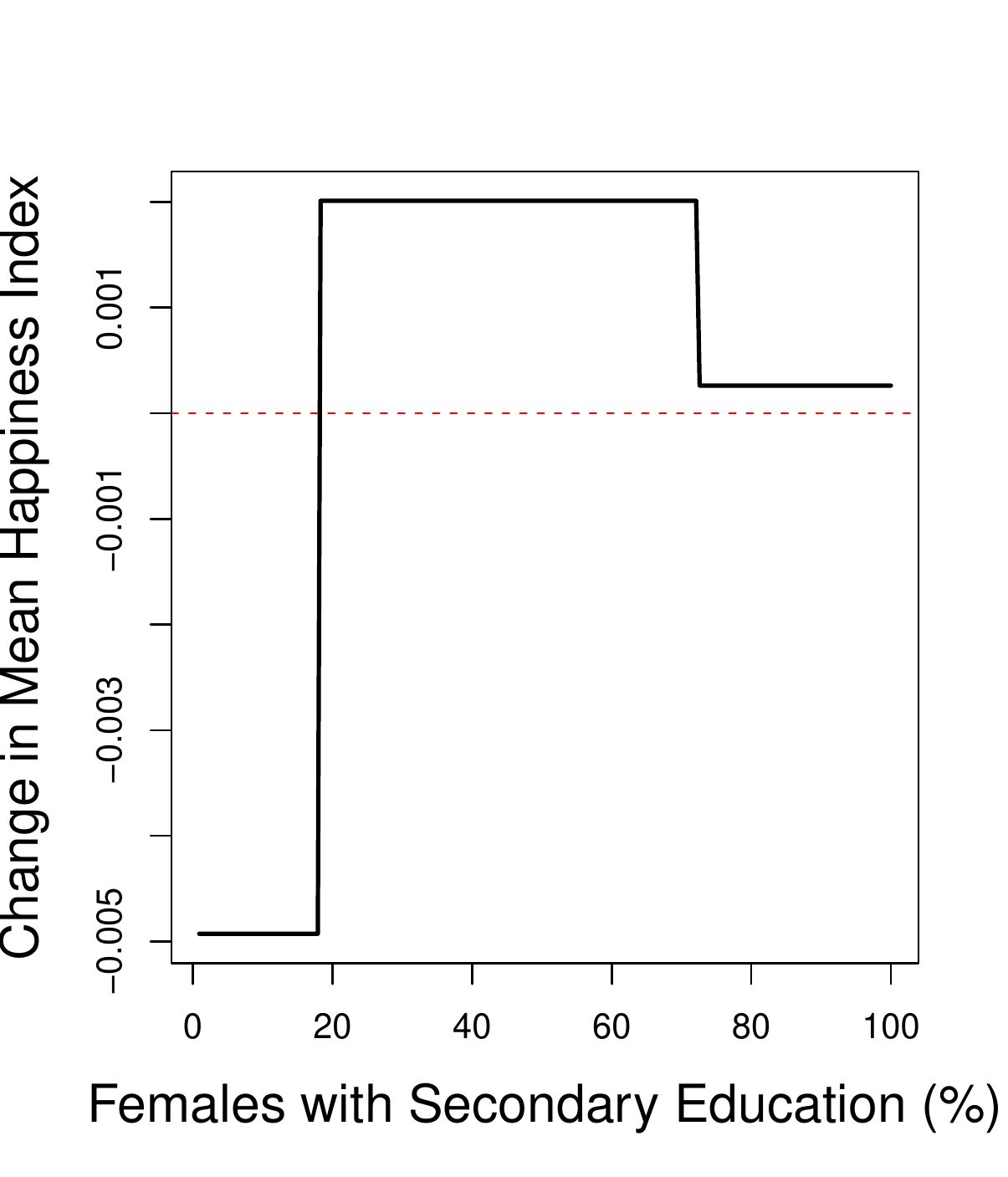}}
	\subfigure{\label{fig:plot7}\includegraphics[width=.264\linewidth]{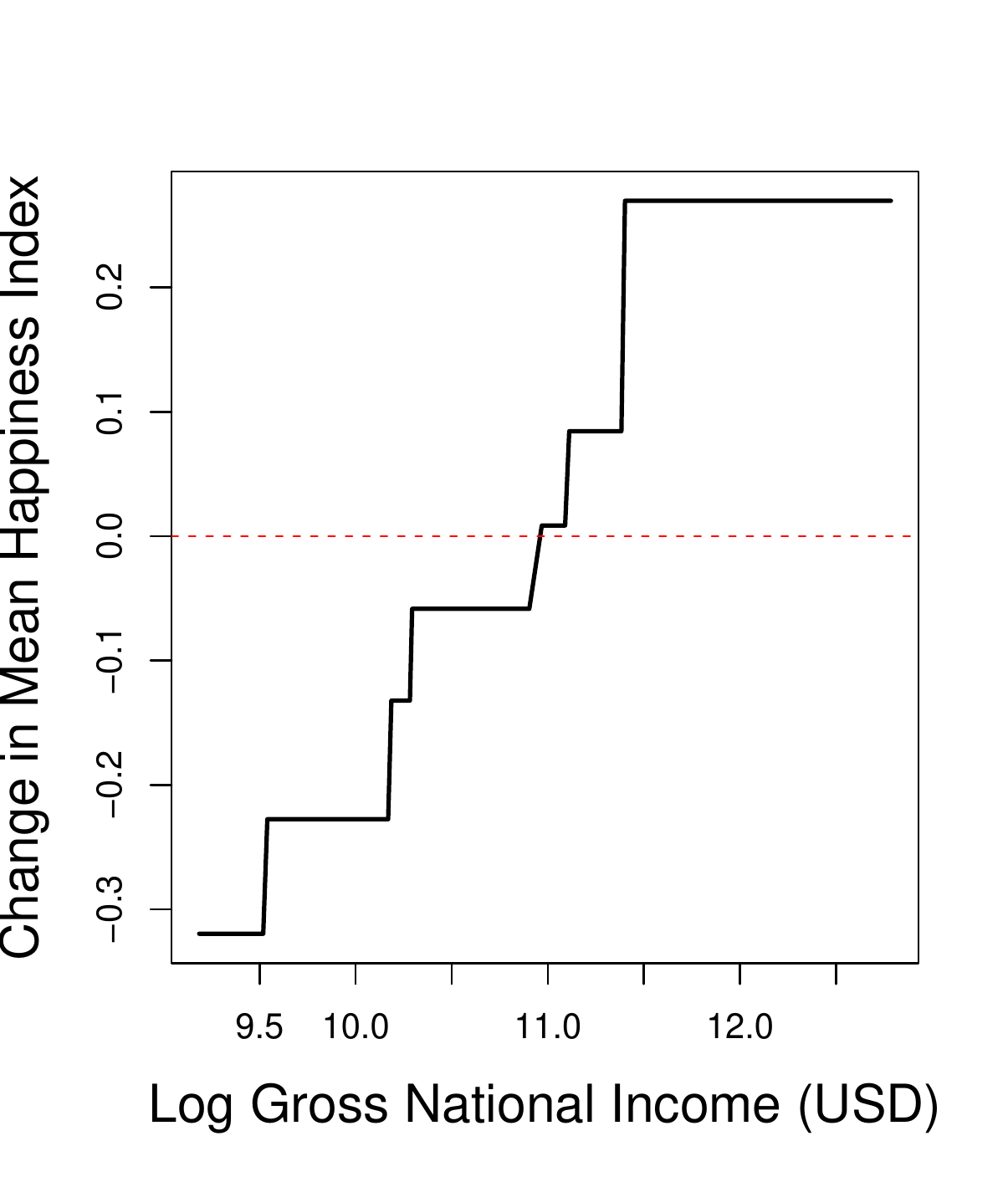}}
	\subfigure{\label{fig:plot9}\includegraphics[width=.264\linewidth]{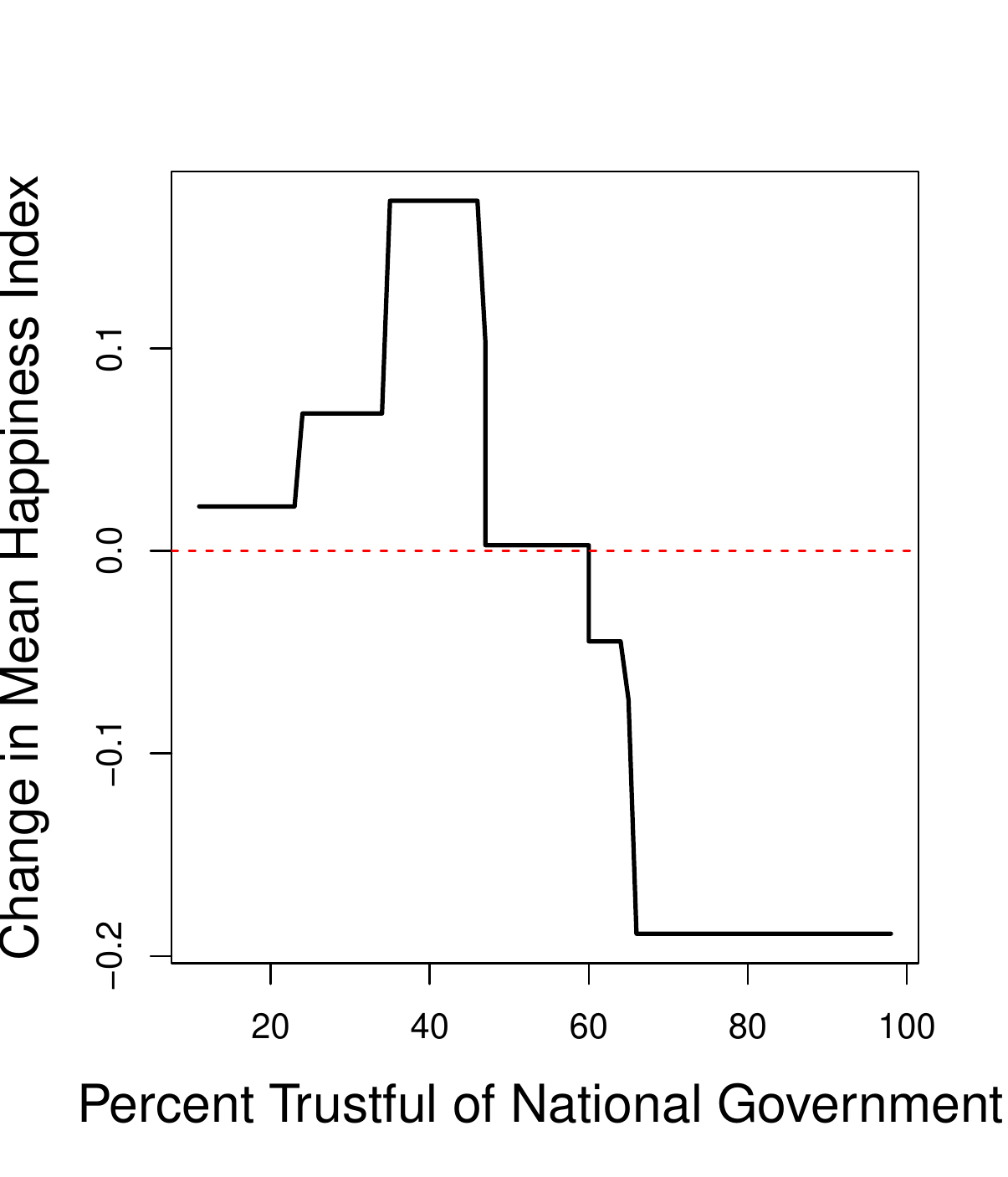}}
	\subfigure{\label{fig:plot10}\includegraphics[width=.264\linewidth]{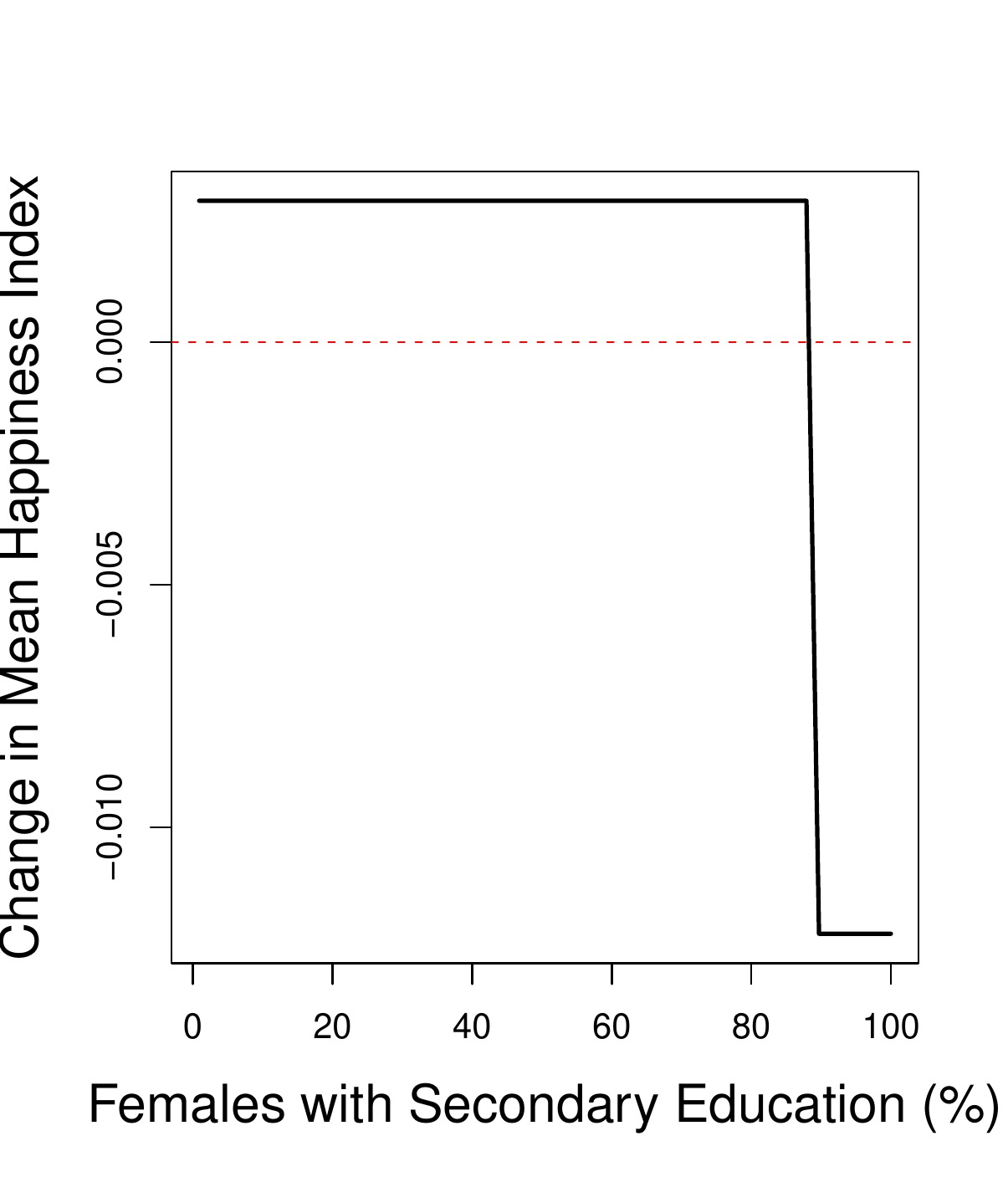}}
	\subfigure{\label{fig:plot11}\includegraphics[width=.264\linewidth]{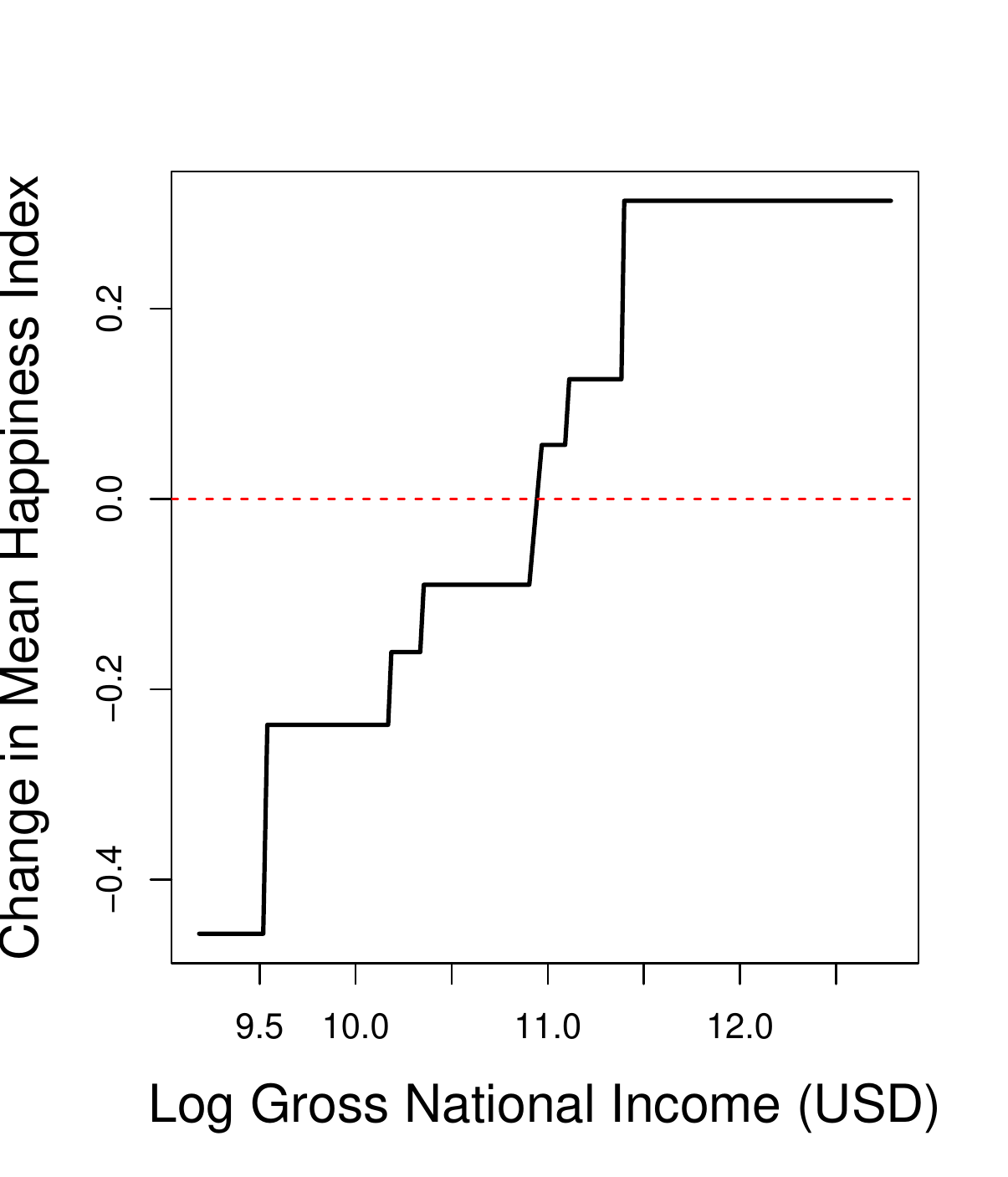}}
	\caption{Conditional associations between country's happiness level and three selected country-level predictors for $\tau=\{0.2,0.5,0.8\}$. The first, second, and third rows correspond to results for $\tau=0.2$, $\tau=0.5$, and $\tau = 0.8$, respectively.}
	\label{fig:happiness}
\end{figure}

We apply the proposed method to investigate the conditional distribution of country-level happiness index at different quantile levels $\tau = \{0.2,0.5,0.8\}$. We implement our proposed method under the Gaussian kernel with $h = \text{max}[0.05,\sqrt{\tau(1-\tau)}\{ \log (p) /n\}^{1/4}]$, and $\lambda$ selected using cross-validation. The estimated fits for three selected covariates are shown in Figure~\ref{fig:happiness}: the first, second, and third rows in Figure~\ref{fig:happiness} correspond to the results for $\tau=0.2$, $\tau=0.5$, and $\tau = 0.8$, respectively. For $\tau=0.5$ the estimated fits are similar to that of the fused lasso additive model presented in \cite{PWS2016}. In particular, we find that an increased in gross national income, up to a certain level, is associated with increased happiness, conditional on the other predictors.  
Moreover, the differences in conditional associations between country's happiness level and both gross national income and percent trustful of national government are negligible for the three quantile levels.  
It is interesting to see that the conditional association between mean happiness index and females with secondary education are different for the three different quantile levels, suggesting that there are potential heterogenous effects.  

\begin{thebibliography}{9}

\bibitem[Beck and Teboulle(2009)]{BT2009}
	{\sc Beck, A.} and {\sc Teboulle,  M.} (2009).
	A fast iterative shrinkage-thresholding algorithm for linear inverse problems.
	{\it SIAM Journal of Imaging Science} {\bf 2} 183--202.
	
\bibitem[Belloni and Chernozhukov(2011)]{BC2011}
	{\sc Belloni, A.} and {\sc Chernozhukov, V.} (2011).
	$\ell_1$-penalized quantile regression in high-dimensional sparse models.
	{\it Annals of Statistics} {\bf 39} 82--130.


\bibitem[B\"uhlmann and van de Geer(2011)]{BVG11}
     {\sc B\"uhlmann, P.} and {\sc van de Geer, S.} (2011).
     {\em Statistics for High-Dimensional Data: Methods, Theory and Applications}. Springer, Heidelberg.


\bibitem[Cantril(1965)]{C1965}
	{\sc Cantril, H.} (1965)
	{\it The Pattern of Human Concerns.}
	Rutgers University Press.
	
\bibitem[Fan and Li(2001)]{FL2001}	
	{\sc Fan, J.} and {\sc Li, R.} (2001).
 	Variable selection via nonconcave regularized likelihood and its oracle properties.
 	{\it Journal of American Statistical Association} {\bf 96} 1348--1360.
	
\bibitem[Fan et~al.(2018)]{FLSF2018}
	{\sc Fan, J., Liu, H., Sun, Q.} and {\sc Zhang, T.} (2018).
	I-LAMM for sparse learning: Simultaneous control of algorithmic complexity and statistical error.
	{\it Annals of Statistics} {\bf 46} 814--841.

\bibitem[Fernandes et~al.(2021)]{FGH2019}
	{\sc Fernandes, M., Guerre, E.} and {\sc Horta, E.} (2021).
	Smoothing quantile regressions.
 	{\it Journal of Business \& Economic Statistics} {\bf 39} 338--357.

\bibitem[Friedma et~al.(2010)]{glmnet}
	{\sc Friedman, J., Hastie, T.} and {\sc Tibshirani, R.} (2010).
	Regularization paths for generalized linear models via coordinate descent.
	{\it Journal of Statistical Software} {\bf 33}(1): 1--22.
	
\bibitem[Helliwell et~al.(2013)]{HLS2013}
	{\sc Helliwell,  J.}, {\sc Layard, R.} and {\sc Sachs,  J.} (2013).
	{\it World Happiness Report 2013.}
	Sustainable Development Solutions Network.

\bibitem[Hunter and Lange(2004)]{HL2004}
	{\sc Hunter, D.\,R.} and {\sc Lange, K.} (2004).
	A tutorial on MM algorithms.
 	{\it The American Statistician} {\bf 58} 30--37.

\bibitem[Hunter and Lange(2000)]{HL2000}
	{\sc Hunter, D.\,R.} and {\sc Lange, K.} (2000).
	Quantile regression via an MM algorithm.
 	{\it Journal of Computational and Graphical Statistics} {\bf 9} 60--77.
 
\bibitem[Gu et~al.(2018)]{Gu2018}	
	{\sc Gu, Y.}, {\sc Fan, J.}, {\sc Kong, L.}, {\sc Ma, S.} and {\sc Zou, H.}  (2018).
 	ADMM for high-dimensional sparse regularized quantile regression.
 	{\it Technometrics} {\bf 60} 319--331.

\bibitem[Hastie et~al.(2015)]{HTW2015}
     {\sc Hastie, T., Tibshirani, R.} and {\sc Wainwright, M.} (2015).
     {\em Statistical Learning with Sparsity: The Lasso and Generalizations}. 
     CRC Press, Boca Raton.

\bibitem[He et~al.(2021)]{HPTZ2020}
	{\sc He, X., Pan, X., Tan, K.\,M.} and {\sc Zhou, W.-X.} (2021).
	Smoothed quantile regression with large-scale inference.
	{\it Journal of Econometrics}, in press.
	
	\bibitem[He et~al.(2022)]{HPTZ2022}
	{\sc He, X., Pan, X., Tan, K.\,M.} and {\sc Zhou, W.-X.} (2022).
	Package ``\texttt{conquer}", version $1.3.0$.
	Reference manual: \href{https://cran.r-project.org/web/packages/conquer/conquer.pdf}{https://cran.r-project.org/web/packages/conquer/conquer.pdf}.
	
\bibitem[Horowitz(1998)]{H1998}
	{\sc Horowitz, J.\,L.} (1998).
	Bootstrap methods for median regression models.
	{\it Econometrica} {\bf 66} 1327--1351.
	
\bibitem[Kato(2011)]{KK2011}
	{\sc Kato,  K.} (2011).
	Group lasso for high dimensional sparse quantile regression models.
	{\it arXiv preprint arXiv:1103.1458.}
	
	\bibitem[Koenker(2005)]{K2005}	
	{\sc Koenker, R.} (2005).
 	{\it Quantile Regression.}
	Cambridge University Press, Cambridge.

\bibitem[Koenker(2022)]{K2022}
	{\sc Koenker, R.} (2022).
	Package ``\texttt{quantreg}", version $5.88$.
	Reference manual: \href{https://cran.r-project.org/web/packages/quantreg/quantreg.pdf}{https://cran.r-project.org/web/packages/quantreg/quantreg.pdf}.


\bibitem[Koenker and Bassett(1978)]{KB1978}	
	{\sc Koenker, R.} and {\sc Bassett, G} (1978).
 	Regression quantiles.
 	{\it Econometrica} {\bf 46} 33-50.

\bibitem[Koenker et~al.(2017)]{KCHP2017}
{\sc Koenker, R., Chernozhukov, V., He, X.} and {\sc Peng, L.} (2017).
{\it Handbook of Quantile Regression.}
CRC Press, Boca Raton, FL.

\bibitem[Koenker and Ng(2005)]{KN2005}
	{\sc Koenker, R.} and {\sc Ng, P.} (2005).
	A Frisch-Newton algorithm for sparse quantile regression.
	{\it Acta Mathematicae Applicatae Sinica} {\bf 21} 225--236.

\bibitem[Lange et~al.(2000)]{LHY2000}
	 {\sc Lange, K.}, {\sc Hunter, D.\,R.} and {\sc Yang, I.} (2000).
	Optimization transfer using surrogate objective functions.
 	{\it Journal of Computational and Graphical Statistics} {\bf 9} 1--59.
		
\bibitem[Li and Zhu(2008)]{LZ2008}
	{\sc Li, Y.} and {Zhu, J.} (2008).
	$\ell_1$-norm quantile regression.
	{\it Journal of Computational and Graphical Statistics} {\bf 17} 163--185.

\bibitem[Pan et~al.(2021)]{PSZ2021}
	{\sc Pan, X.}, {\sc Sun, Q.} and {\sc Zhou, W.-X.} (2021).
	Iteratively reweighted $\ell_1$-penalized robust regression.
	{\it Electronic Journal of Statistics} {\bf 25} 3287--3348.

\bibitem[Peng and Wang(2015)]{PW2015}
	{\sc Peng, B.} and {Wang, L.} (2015).
	An iterative coordinate descent algorithm for high-dimensional nonconvex penalized quantile regression.
	{\it Journal of Computational and Graphical Statistics} {\bf 24} 676--694.

\bibitem[Petersen and Witten(2019)]{PW2019}
	{\sc Petersen, A.} and {\sc Witten, D.} (2019).
	 Data-adaptive additive modeling.
 	{\it Statistics in Medicine} {\bf 38} 583--600.
	
\bibitem[Petersen et~al.(2016)]{PWS2016}
	{\sc Petersen, A.}, {\sc Witten, D.} and {\sc Simon, N.} (2016).
	 Fused lasso additive model.
 	{\it Journal of Computational and Graphical Statistics} {\bf 25} 1005--1025.

	
\bibitem[Portnoy and Koenker(1997)]{PK1997}
	{\sc Portnoy, S.} and {\sc Koenker, R.} (1997).
	The Gaussian hare and the Laplacian tortoise: Computability of squared-error versus absolute-error estimators.
	{\it Statistical Science} {\bf 12} 279--300.


\bibitem[Sadhanala and Tibshirani(2019)]{ST2019}	
	{\sc Sadhanala, V.} and {\sc Tibshirani, R.} (2019).
	Additive models with trend filtering.
 	{\it Annals of Statistics} {\bf 47} 3032--3068.
		
\bibitem[Sherwood and Maidman(2020)]{SM2020}
	{\sc Sherwood, B.} and {\sc Maidman, A.} (2020).
	Package ``\texttt{rqPen}", version $2.2.2$.
	Reference manual: \href{https://cran.r-project.org/web/packages/rqPen/rqPen.pdf}{https://cran.r-project.org/web/packages/rqPen/rqPen.pdf}.
	

	
\bibitem[Simon et~al.(2013)]{SFHT2013}
	{\sc Simon, N.}, {\sc Friedman, J.}, {Hastie, T.} and {Tibshirani, R.} (2013).
	A sparse-group lasso.
	{\it Journal of Computational and Graphical Statistics} {\bf 22}(2): 231--245.
	
\bibitem[Tan et~al.(2022)]{TWZ2021}
	{\sc Tan, K.  M.}, {\sc Wang, L.} and {\sc Zhou, W.-X.} (2022).
	High-dimensional quantile regression: Convolution smoothing and concave regularization.
	{\it Journal of the Royal Statistical Society: Series B} {\bf 84}(1): 205--233.


\bibitem[Tibshirani(1996)]{Tibs1996}	
	{\sc Tibshirani, R.} (1996).
 	Regression shrinkage and selection via the lasso.
 	{\it Journal of the Royal Statistical Society: Series B} {\bf 58} 267--288.

\bibitem[Tibshirani(2014)]{T2014}
	{\sc Tibshirani, R.} (2014).
	 Adaptive piecewise polynomial estimation via trend filtering.
 	{\it Annals of Statistics} {\bf 42} 285--323.
 	
 \bibitem[Tibshirani et~al.(2005)]{TSRZK2005}
 	{\sc Tibshirani, R.} and {\sc Saunders, M.}, {\sc Rosset, S.}, {\sc Zhu, J.} and {\sc Knight, K.} (2005).
 	Sparsity and smoothness via the fused lasso.
 	{\it Journal of the Royal Statistical Society: Series B} {\bf 67}  91--108.

\bibitem[Tibshirani and Taylor(2011)]{TT2011}
	{\sc Tibshirani, R.} and {\sc Taylor, J.} (2011).
	 The solution path of the generalized lasso. 
 	{\it Annals of Statistics} {\bf 39} 1335--1371.

\bibitem[United Nations Development Programme(2012)]{UNDP2012}
	{\sc United Nations Development Programme.} (2012).
	Human Development Indicators. 
	{\it United Nations Publications.}

\bibitem[Wainwright(2019)]{W2019}	
	{\sc Wainwright, M.\,J.} (2019).
 	{\it High-Dimensional Statistics: A Non-Asymptotic Viewpoint.}
 	{Cambridge University Press, Cambridge.}
	
 	
\bibitem[Wang et~al.(2012)]{WWL2012}	
	{\sc Wang, L.}, {\sc Wu, Y.} and {\sc Li, R.} (2012).
 	Quantile regression for analyzing heterogeneity in ultra-high dimension.
 	{\it Journal of American Statistical Association} {\bf 107} 214--222.
	
\bibitem[World Bank Group(2012)]{WBG2012}
	{\sc World Bank Group.} (2012).
	World Development Indicators 2012. 
	{\it World Bank Publications.}

\bibitem[Wu and Witten(2019)]{WW2019}
	{\sc Wu, J.} and {\sc Witten, D.} (2019).
	 Flexible and interpretable models for survival data.
 	{\it Journal of Computational and Graphical Statistics} {\bf 28} 954--966.
	
	
\bibitem[Yang and Zou(2015)]{YZ2015}	
	{\sc Yang, Y.} and {\sc Zou, H.} (2015).
 	A fast unified algorithm for solving group-lasso penalized learning problems. 
 	{\it Statistics and Computing} {\bf 25} 1129--1141.

\bibitem[Yi and Huang(2017)]{YH2017}
	{\sc Yi, C.} and {\sc Huang, J.} (2017).
	Semismooth Newton coordinate descent algorithm for elastic-net penalized Huber regression and quantile regression.
	{\it Journal of Computational and Graphical Statistics} {\bf 26} 547--557.

\bibitem[Yu et~al.(2017)]{Yu2017}	
	{\sc Yu, L.}, {\sc Lin, N.} and {\sc Wang, L.} (2017).
 	A parallel algorithm for large-scale nonconvex penalized quantile regression.
 	{\it Journal of Computational and Graphical Statistics} {\bf 26} 935--939.
	
\bibitem[Yuan and Lin(2006)]{YL2006}
	{\sc Yuan, M.} and {\sc Lin, Y.} (2006).
	Model selection and estimation in regression with grouped variables.
	{\it Journal of the Royal Statistical Society: Series B} {\bf 68} 49--67.

\bibitem[Zhang(2010)]{ZMCP2010}	
	{\sc Zhang, C.-H.} (2010).
 	Nearly unbiased variable selection under minimax concave penalty.
 	{\it Annals of Statistics} {\bf 38} 894--942.

\bibitem[Zheng et~al.(2015)]{ZPH2015}	
	{\sc Zheng, Q.}, {\sc Peng, L.} and {\sc He, X.} (2015).
	Globally adaptive quantile regression with ultra-high dimensional data.
 	{\it Annals of Statistics} {\bf 43} 2225--2258.
		
\bibitem[Zou and Hastie(2005)]{ZH2005}
	{\sc Zou, H.} and {\sc Hastie, T.} (2005).
	Regularization and variable selection via the elastic net.
	{\it Journal of the Royal Statistical Society: Series B} {\bf 67} 301--320.


\end{thebibliography}
\end{document}